\documentclass[a4paper,11pt]{article}%
\usepackage{amsmath}
\usepackage{cite}
\usepackage{graphicx}%
\usepackage{amsfonts}%
\usepackage{amssymb}
\begin{document}
\begin{flushright}
UCL-IPT-05-08

17 August 2005
\end{flushright}

\vskip  4 true cm

\begin{center}
{\Large RADIATIVE KAON DECAYS AND THE PENGUIN\medskip}

{\Large CONTRIBUTION TO THE
\Large{$\Delta I = 1/2$}%
\ RULE}\\[45pt]\textsc{Jean-Marc G\'{e}rard},${}^{1}$ \textsc{Christopher
Smith},${}^{2}$ and \textsc{St\'{e}phanie Trine}${}^{2}$\\[30pt]${}^{1}%
~$\textsl{Institut de Physique Th\'{e}orique, Universit\'{e} catholique de
Louvain, }\\[0pt]\textsl{Chemin du Cyclotron 2, B-1348 Louvain-la-Neuve,
Belgium} \\[6pt]${}^{2}~$\textsl{INFN, Laboratori Nazionali di Frascati, Via
E. Fermi 40, I-00044 Frascati, Italy} \\[45pt]\textbf{Abstract}
\end{center}

A consistent census of penguins in the $\Delta I=1/2$ rule is taken from the
$\eta_{0}$ pole contribution to the radiative $K_{L}\rightarrow\gamma\gamma$,
$K_{S}\rightarrow\pi^{0}\gamma\gamma$ and $K^{+}\rightarrow\pi^{+}\gamma
\gamma$ decay modes. We briefly comment on its impact for $K_{L}\rightarrow
\pi^{0}\pi^{0}\gamma\gamma$, $K_{L}\rightarrow\pi^{+}\pi^{-}\gamma$ and check
its compatibility with the $K_{L}-K_{S}$ mass difference and the CP violating
$\varepsilon^{\prime}/\varepsilon$ parameter.

\setcounter{footnote}{0}
\pagebreak 

\section{Introduction}

A precise quantitative understanding of hadronic kaon decays has, up to now,
been upset by the non-perturbative nature of strong interactions at low
energy. Though qualitatively, there is little doubt that the $\Delta I=1/2$
rule is a pure QCD effect, the genuine mechanism still eludes us. In
particular, the relative strength of penguin and current-current operators has
been debated for years. This has in turn impeded theorists from making use of
the remarkable precision achieved in $\varepsilon^{\prime}/\varepsilon$ measurements.

In the present work, this question is addressed from a phenomenological point
of view. The relative contribution of penguin and current-current operators in
the $\Delta I=1/2$ rule is known to be accessible from anomaly-driven
radiative $K$ decays, i.e. decays occurring through $\pi,\eta,\eta^{\prime}$
poles. Still, a thorough investigation in the context of large-$N_{c}$ Chiral
Perturbation Theory (ChPT) is called for. In particular, we will see that
$U(3)$ ChPT holds the key to understand, and go beyond, the well-known
vanishing of the $SU(3)$ $K_{L}\rightarrow\gamma\gamma$ amplitude at lowest
order. Besides, the experimental situation has been improved and is expected
to continue doing so for a number of radiative $K$ decays. It is thus
appropriate to review their theoretical treatment, especially regarding
$\eta_{0}$ effects, and to guide experimentalists towards suitable observables
to probe the $\Delta I=1/2$ rule.

The paper is organized as follows. In the first section, the large-$N_{c}$
formalism is recalled, with emphasis on the $U(3)$ ChPT weak operator basis.
In particular, the strategy to extract the penguin fraction from phenomenology
is exposed. The next few sections deal with radiative $K$ decays. We first
review in great details the $K_{L}\rightarrow\gamma\gamma$ mode, and develop
the phenomenological tools to deal with $\eta-\eta^{\prime}$ mixing to be used
in the rest of the paper. Then we go on with the $K_{S}\rightarrow\pi
^{0}\gamma\gamma$ decay and the pole contribution to the $K^{+}\rightarrow
\pi^{+}\gamma\gamma$ one, which are both quite sensitive to $\eta-\eta
^{\prime}$ effects. The $K_{L}\rightarrow\pi^{0}\pi^{0}\gamma\gamma$ and
$K_{L}\rightarrow\pi^{+}\pi^{-}\gamma$ modes are only briefly analyzed as no
new information can be obtained from them. Finally, the implications for the
hadronic observables $\Delta M_{LS}$ and $\varepsilon^{\prime}/\varepsilon$
are investigated. Our results are summarized in the conclusion.

In the appendix, the complete analysis of $K_{L}\rightarrow\gamma\gamma$ in
$SU(3)$ at $\mathcal{O}(p^{6})$ is presented, some aspects of which are used
throughout the paper to investigate the reduction of $U(3)$ amplitudes to
$SU(3)$ ones.

\section{Theoretical framework}

Our starting point is the QCD-induced $\Delta S=1$ effective weak Hamiltonian
below the charm mass scale \cite{WeakHam,BuchallaBL96}:%
\begin{equation}
\mathcal{H}_{eff}^{\Delta S=1}\left(  \mu<m_{c}\right)  =\frac{G_{F}}{\sqrt
{2}}V_{ud}V_{us}^{\ast}\sum_{i=1}^{6}\left[  z_{i}\left(  \mu\right)
-\frac{V_{td}V_{ts}^{\ast}}{V_{ud}V_{us}^{\ast}}\,\,y_{i}\left(  \mu\right)
\right]  Q_{i} \label{EffWeakHam}%
\end{equation}
with the familiar current-current $(i=1,2)$ and penguin $(i=3,...,6)$
operators%
\begin{equation}%
\begin{array}
[c]{ll}%
Q_{1}=4\left(  \bar{s}_{L}\gamma_{\alpha}d_{L}\right)  \left(  \bar{u}_{L}%
\gamma^{\alpha}u_{L}\right)  ,\smallskip & Q_{2}=4\left(  \bar{s}_{L}%
\gamma_{\alpha}u_{L}\right)  \left(  \bar{u}_{L}\gamma^{\alpha}d_{L}\right)
,\\
Q_{3}=4\left(  \bar{s}_{L}\gamma_{\alpha}d_{L}\right)  \left(  \bar{q}%
_{L}\gamma^{\alpha}q_{L}\right)  ,\smallskip & Q_{4}=4\left(  \bar{s}%
_{L}\gamma_{\alpha}q_{L}\right)  \left(  \overline{q}_{L}\gamma^{\alpha}%
d_{L}\right)  ,\\
Q_{5}=4\left(  \bar{s}_{L}\gamma_{\alpha}d_{L}\right)  \left(  \bar{q}%
_{R}\gamma^{\alpha}q_{R}\right)  ,\quad & Q_{6}=-8\left(  \bar{s}_{L}%
q_{R}\right)  \left(  \bar{q}_{R}d_{L}\right)  ,
\end{array}
\label{QiOp}%
\end{equation}
after Fierz reorderings. In our notations, $q_{L}^{R}\equiv\frac{1}{2}%
(1\pm\gamma_{5})q$ and the light flavors $q=u,d,s$ are summed over. The
connection between Eq.(\ref{EffWeakHam}) and kaon phenomenology requires of
course additional tools, QCD perturbation theory being helpless to estimate
the $Q_{i}$ hadronic matrix elements. We will make use of chiral Lagrangian
techniques, as we now describe.

\subsection{U(3) chiral representation of $\Delta S=1$ weak operators}

As is well-known, the above operators are not all independent. Indeed, we have
the identity%
\begin{equation}
Q_{2}-Q_{1}=Q_{4}-Q_{3}.
\end{equation}
Besides, $Q_{4}$ and $Q_{6}$ have the same color and flavor structures. The
chiral realization of the effective Hamiltonian $\mathcal{H}_{eff}^{\Delta
S=1}$ must thus allow for an explicit representation of the left-handed
flavor-singlet current in $Q_{3}$ if we want to be in a position to
disentangle $Q_{1,2}$ and $Q_{6}$ contributions to the $\Delta I=1/2$ kaon
decay amplitudes. For this reason, we will perform the hadronization of
Eq.(\ref{EffWeakHam}) in the $U(3)$ chiral expansion, with an explicit
flavor-singlet degree of freedom $\eta_{0}$, rather than in $SU(3)$.

The extension of $SU(3)$ to $U(3)$ proceeds through the large-$N_{c}$ limit,
with $N_{c}$ the number of QCD colors. Considering $N_{c}$ as large is the key
to a consistent disposal of the QCD $U(1)_{A}$ anomaly \cite{Witten79}. The
spontaneous symmetry breaking $U(3)_{L}\times U(3)_{R}\rightarrow U(3)_{V}$
gives then rise to a nonet of pseudoscalar Goldstone bosons, which are written%
\begin{equation}
U\equiv\exp i\frac{\sqrt{2}}{F}\left(
\begin{array}
[c]{ccc}%
\dfrac{\pi^{0}}{\sqrt{2}}+\dfrac{\eta_{8}}{\sqrt{6}}+\dfrac{\eta_{0}}{\sqrt
{3}} & \pi^{+} & K^{+}\\
\pi^{-} & -\dfrac{\pi^{0}}{\sqrt{2}}+\dfrac{\eta_{8}}{\sqrt{6}}+\dfrac
{\eta_{0}}{\sqrt{3}} & K^{0}\\
K^{-} & \bar{K}^{0} & -\dfrac{2\eta_{8}}{\sqrt{6}}+\dfrac{\eta_{0}}{\sqrt{3}}%
\end{array}
\right)
\end{equation}
in the exponential parametrization. The anomalous breaking of the $U(1)_{A}$
symmetry is reintroduced through $1/N_{c}$ corrections. In particular, the
leading nonlinear Lagrangian reads \cite{NLLagr}%
\begin{equation}
\mathcal{L}_{S}^{(p^{2},\infty)+(p^{0},1/N_{c})}=\dfrac{F^{2}}{4}%
\langle\partial_{\mu}U\partial^{\mu}U^{\dagger}\rangle+\dfrac{F^{2}}{4}%
\langle\chi U^{\dagger}+U\chi^{\dagger}\rangle+\dfrac{F^{2}}{16N_{c}}m_{0}%
^{2}\langle\ln U-\ln U^{\dagger}\rangle^{2} \label{LagrNL}%
\end{equation}
where $\left\langle {}\right\rangle $ denotes a trace over flavors, the
external source $\chi$ is frozen at $\chi=rM$ to account for meson masses,
$M=diag(m_{u},m_{d},m_{s})$ is the light quark mass matrix and $m_{0}$
represents the anomalous part of the $\eta_{0}$ mass. The ChPT
symmetry-breaking scale $F$ is identified with the neutral pion decay constant
$F_{\pi}=92.4$ MeV at this order. Note that the leading $SU(3)$ chiral
Lagrangian is recovered in the limit $m_{0}\rightarrow\infty$, when the
$\eta_{0}$ decouples.

The effective Hamiltonian (\ref{EffWeakHam}) contains both $(8_{L},1_{R})$ and
$(27_{L},1_{R})$ representations of the chiral group $U(3)_{L}\times U(3)_{R}%
$. At leading order in the momentum expansion, four operators built out of $U$
and $\chi$ have either of these transformation properties and at most one
factorized $\eta_{0}$ field\footnote{Operators with two or more $\eta_{0}$
factors have negligible effects on the processes we will consider due to
isospin, loop and/or $1/N_{c}$ suppressions.}:%
\begin{equation}%
\begin{tabular}
[c]{l}%
$Q_{8}=4\left(  L_{\mu}L^{\mu}\right)  _{23},\quad Q_{8}^{s}=4\left(  L_{\mu
}\right)  _{23}\left\langle L^{\mu}\right\rangle ,\quad Q_{8}^{m}=F^{4}\left(
\chi U^{\dagger}+U\chi^{\dagger}\right)  _{23},\smallskip\smallskip$\\
$Q_{27}=4\left[  \left(  L_{\mu}\right)  _{23}\left(  L^{\mu}\right)
_{11}+\frac{2}{3}\left(  L_{\mu}\right)  _{13}\left(  L^{\mu}\right)
_{21}-\frac{1}{3}\left(  L_{\mu}\right)  _{23}\left\langle L^{\mu
}\right\rangle \right]  ,$%
\end{tabular}
\label{Q8Q8mQ27}%
\end{equation}
where the matrix $L_{\mu}$ collects the chiral realizations of the left-handed
currents from Eq.(\ref{LagrNL}):%
\begin{equation}
\bar{q}_{L}^{k}\gamma_{\mu}q_{L}^{l}\rightarrow i\frac{F^{2}}{2}\left(
\partial_{\mu}UU^{\dagger}\right)  ^{lk}\equiv\left(  L_{\mu}\right)
^{lk}\quad(q^{l,k}=u,d,s). \label{Lmu}%
\end{equation}
$Q_{8}$, $Q_{8}^{m}$ and $Q_{27}$ are the trivial $U(3)$ generalizations of
the usual $SU(3)$ octet and 27 operators while%
\begin{equation}
Q_{8}^{s}\sim\left(  L_{\mu}\right)  _{23}\,\partial^{\mu}\eta_{0}
\label{Q8sStructure}%
\end{equation}
is peculiar to the $U(3)$ framework. Note that the introduction of one extra
$\langle\ln U-\ln U^{\dagger}\rangle\sim\eta_{0}$ factor in $Q_{8}$,
$Q_{8}^{m}$ or $Q_{27}$ would violate $CPS$ invariance. Besides, other
structures like $(\chi U^{\dagger}-U\chi^{\dagger})_{23}\langle\ln U-\ln
U^{\dagger}\rangle$ can be brought back to $Q_{8}^{s}$ through the use of the
classical equations of motion. For our purpose, the complete low-energy
realization of $\mathcal{H}_{eff}^{\Delta S=1}$ at $\mathcal{O}(p^{2})$ is
thus given by%
\begin{equation}
\mathcal{H}_{W}^{\Delta S=1}=G_{8}Q_{8}+G_{8}^{s}Q_{8}^{s}+G_{8}^{m}Q_{8}%
^{m}+G_{27}Q_{27}. \label{LagrWspurion}%
\end{equation}
Yet the weak mass term $Q_{8}^{m}$ is known not to contribute to hadronic
observables at leading order, while its higher order effects can be absorbed
into a redefinition of the weak counterterms \cite{KamborMW90,WeakMassTerm}.
We will thus disregard it from now. Later, we will check explicitly that the
same holds true in respect of radiative kaon decays.

Let us now make a closer connection with the current-current and penguin
operators of Eq.(\ref{QiOp}).

\subsection{QCD-inspired alternative basis}

As a first step towards an alternative QCD-inspired formulation of
Eq.(\ref{LagrWspurion}), we consider the effective Hamiltonian
(\ref{EffWeakHam}) in the factorization approximation.

In that limit, the chiral realization $\hat{Q}_{i}$ of the $(V-A)\otimes(V\pm
A)$ $Q_{i}$ operators is straightforward. It simply follows from the
hadronization of the left- and right-handed currents in Eq.(\ref{Lmu}) (with
$U\leftrightarrow U^{\dagger}$ for the right-handed currents). This gives:%
\begin{equation}%
\begin{tabular}
[c]{ll}%
$\hat{Q}_{1}=4\left(  L_{\mu}\right)  _{23}\left(  L^{\mu}\right)
_{11},\smallskip$ & $\hat{Q}_{2}=4\left(  L_{\mu}\right)  _{13}\left(  L^{\mu
}\right)  _{21},$\\
$\hat{Q}_{3}=-Q_{5}=4\left(  L_{\mu}\right)  _{23}\left\langle L^{\mu
}\right\rangle ,\quad$ & $\hat{Q}_{4}=4\left(  L_{\mu}L^{\mu}\right)  _{23}. $%
\end{tabular}
\end{equation}
For the $\hat{Q}_{6}$ penguin operator, the situation is more delicate as the
constraint $UU^{\dagger}=1$ forces us to go beyond Eq.(\ref{LagrNL}) to
hadronize the quark densities \cite{ChivukulaFG86,BardeenBG86}. From the
$\mathcal{O}(p^{4})$ strong Lagrangian in the large-$N_{c}$ limit%
\begin{equation}
\mathcal{L}_{S}^{(p^{4},\infty)}\ni\langle\partial_{\mu}U\partial^{\mu
}U^{\dagger}(\chi U^{\dagger}+U\chi^{\dagger})\rangle,\,\langle\chi\square
U^{\dagger}+\square U\chi^{\dagger}\rangle,\,\langle(\chi U^{\dagger}\pm
U\chi^{\dagger})^{2}\rangle, \label{LagrNLp4}%
\end{equation}
two structures%
\begin{equation}
\hat{Q}_{6}=4\left(  L_{\mu}L^{\mu}\right)  _{23}%
\end{equation}
(identical with $\hat{Q}_{4}$) and $Q_{8}^{m}$ emerge.

Corrections with respect to the factorization approximation consist in meson
exchanges between currents or densities. Such $1/N_{c}$ loop effects, despite
formally suppressed, are important due to the quadratic dependence on the
physical cut-off for a truncated chiral Lagrangian describing the low-energy
strong interactions of massless pseudoscalars in terms of the scale
$F$\cite{BardeenBG87}. Accordingly, no new structure will be induced at
$\mathcal{O}(p^{2})$ since the other scale parameter $m_{0}$ in
Eq.(\ref{LagrNL}) has no effect on this fast operator
evolution\cite{FateloG95}. We thus reach the ansatz:%
\begin{equation}
\mathcal{H}_{W}^{\Delta S=1}=G_{W}\sum_{i=1}^{6}x_{i}\hat{Q}_{i}.
\label{LagrWqcd}%
\end{equation}

The overall constant $G_{W}$ and the weights $x_{i}$ are not fixed by symmetry
arguments. Let us set $G_{W}\equiv G_{F}V_{ud}V_{us}^{\ast}/\sqrt
{2}=1.77\times10^{-12}$ MeV$^{-2}$ so that in the factorization approximation
(neglecting $V_{td}V_{ts}^{\ast}$):%
\begin{equation}
x_{i}\overset{FACT.}{=}z_{i}\left(  \mu_{fact}\right)  \;\;\;(i\neq
6),\;\;\;\;\;\;\;x_{6}\overset{FACT.}{=}-r^{2}\;\frac{F_{K}/F_{\pi}-1}%
{m_{K}^{2}-m_{\pi}^{2}}\;z_{6}\left(  \mu_{fact}\right)  , \label{EqNorm}%
\end{equation}
with $\mu_{fact}$ around 1 GeV. The difference $F_{K}/F_{\pi}-1$ is generated
by a linear combination of the first two terms of Eq.(\ref{LagrNLp4}).

The actual $x_{i}$ values differ from those given in Eq.(\ref{EqNorm}) by
long-distance strong interaction effects. Following the principles of chiral
expansions, we wish to extract them from phenomenology. Yet only three
combinations of $x_{i}$ are accessible as the corresponding operators can all
be expressed in terms of $Q_{8}$, $Q_{8}^{s}$ and $Q_{27}$. Identifying
Eqs.(\ref{LagrWspurion}) and (\ref{LagrWqcd}), we obtain:%
\begin{align}
G_{8}/G_{W}  &  =-\frac{2}{5}x_{1}+\frac{3}{5}x_{2}+x_{4}+x_{6}\nonumber\\
G_{8}^{s}/G_{W}  &  =\frac{3}{5}x_{1}-\frac{2}{5}x_{2}+x_{3}-x_{5}%
\label{ChangeBasis}\\
G_{27}/G_{W}  &  =\frac{3}{5}\left(  x_{1}+x_{2}\right)  .\nonumber
\end{align}
In order to invert the above system, some extra theoretical assumptions are
needed. Corrections with respect to the factorization limit are likely to be
important in the case of $x_{6}$ due to the presence of the factor
$r=2m_{K}^{2}/(m_{s}+m_{d})$ in Eq.(\ref{EqNorm}). We thus have to keep it.
However, the situation is different for $x_{3}$, $x_{4}$ and $x_{5}$, whose
small factorized values suggest marginal effects. Neglecting $\hat{Q}_{3}$,
$\hat{Q}_{4}$ and $\hat{Q}_{5}$, the QCD-inspired set $(\hat{Q}_{1},\hat
{Q}_{2},\hat{Q}_{6})$ becomes equivalent to the basis $(Q_{8},Q_{8}^{s}%
,Q_{27})$. Notice again that, had we worked in the $SU(3)$ framework, the
disappearance of the second octet $Q_{8}^{s}$ would have led to a linear
relation between $\hat{Q}_{1}$, $\hat{Q}_{2}$ and $\hat{Q}_{6}$, rendering
thus impossible the distinction between current-current and penguin operators
from phenomenology.

\subsection{A first look at the weak couplings}

Our goal is to extract $x_{1}$, $x_{2}$ and $x_{6}$. A first piece of
information can readily be obtained from the analysis of the $K\rightarrow
\pi\pi$ decays. In the isospin limit ($m_{u}=m_{d}$, no electromagnetic
corrections), these can be parametrized in terms of two isospin amplitudes,
$A_{0}$ $(\Delta I=1/2)$ and $A_{2}$ $(\Delta I=3/2)$:%
\begin{align}
\mathcal{A}\left(  K^{0}\rightarrow\pi^{+}\pi^{-}\right)   &  =A_{0}%
e^{i\delta}+\dfrac{1}{\sqrt{2}}A_{2},\nonumber\\
\mathcal{A}\left(  K^{0}\rightarrow\pi^{0}\pi^{0}\right)   &  =A_{0}%
e^{i\delta}-\sqrt{2}A_{2},\smallskip\label{Eq16}\\
\mathcal{A}\left(  K^{+}\rightarrow\pi^{+}\pi^{0}\right)   &  =\dfrac{3}%
{2}A_{2},\nonumber
\end{align}
with $\delta\equiv\delta_{0}-\delta_{2}\simeq\pi/4$, the final state
interaction phase shift. In our conventions, $A_{2}$ is real and positive in
the limit of $CP$ conservation, adopted from now on. The $\Delta I=1/2$ piece
receives contributions from both the $Q_{8}$ and $Q_{27}$ operators (none from
$Q_{8}^{s}$ in the isospin limit), while the $\Delta I=3/2$ piece is only
brought about by $Q_{27}$. In terms of the $x_{i}$'s, this gives:%
\begin{align}
A_{0}  &  =\sqrt{2}F\left(  G_{8}+\frac{1}{9}G_{27}\right)  (m_{K}^{2}-m_{\pi
}^{2})=\frac{\sqrt{2}}{3}F\,G_{W}\,\left(  -x_{1}+2x_{2}+3x_{6}\right)
(m_{K}^{2}-m_{\pi}^{2}),\label{KtoPiPiA0}\\
A_{2}  &  =\frac{10}{9}F\,G_{27}\,(m_{K}^{2}-m_{\pi}^{2})=\frac{2}{3}%
F\,G_{W}\,\left(  x_{1}+x_{2}\right)  (m_{K}^{2}-m_{\pi}^{2}),
\label{KtoPiPiA2}%
\end{align}
with $x_{4}$ neglected (as indicated above) and no contribution from $x_{3,5}
$. Experimentally, $\omega^{-1}\equiv A_{0}/A_{2}=22.2$, which is the very
statement of the $\Delta I=1/2$ rule. This large factor implies a dominance of
$G_{8}$ over $G_{27}$:
\begin{equation}
G_{8}=9.1\times10^{-12}\,\text{MeV}^{-2},\quad G_{27}=5.3\times10^{-13}%
\,\text{MeV}^{-2}. \label{G8G27}%
\end{equation}
However, this does not give any information yet on the relative sizes of the
$x_{i}$'s, and thus on the current-current and penguin fractions in $A_{0}$%
\begin{equation}
\mathcal{F}_{CC}\equiv\frac{-x_{1}+2x_{2}}{-x_{1}+2x_{2}+3x_{6}}%
\quad\text{and}\quad\mathcal{F}_{P}\equiv\frac{3x_{6}}{-x_{1}+2x_{2}+3x_{6}%
}\;. \label{CCPFrac}%
\end{equation}%
\begin{figure}
[t]
\begin{center}
\includegraphics[
height=2.1274in,
width=3.6997in
]%
{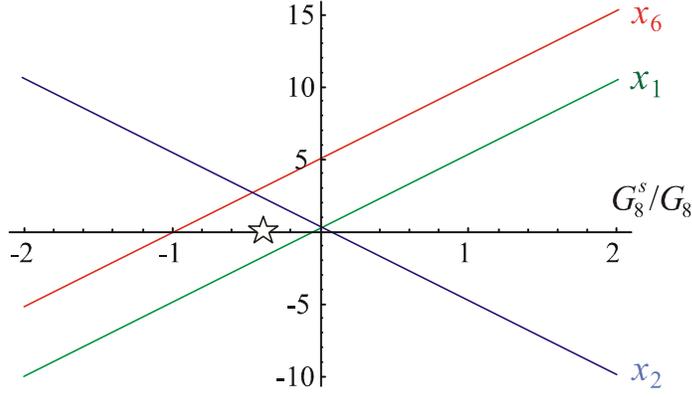}%
\caption{$x_{1}$, $x_{2}$ and $x_{6}$ as a function of $G_{8}^{s}/G_{8}$
($G_{8}$, $G_{27}$ fixed by $K\rightarrow\pi\pi$). The star indicates the
QCD-inspired estimate of Eq.(\ref{QCDinspG8s}).}%
\end{center}
\end{figure}

The missing piece of information is of course $G_{8}^{s}$. Before proceeding
to its phenomenological determination, let us see what comes out from
additional QCD-inspired assumptions. The values of $x_{1}$, $x_{2}$ and
$x_{6}$ corresponding to a given ratio $G_{8}^{s}/G_{8}$ after imposition of
the constraints (\ref{G8G27}) on the system (\ref{ChangeBasis}) are depicted
in Fig.1. In the case of $x_{1}$ and $x_{2}$, strong interaction effects are
not expected to change the signs of the factorized values (\ref{EqNorm}),
i.e., we should keep $x_{1}<0$ and $x_{2}>0$. This favours a negative ratio
$G_{8}^{s}/G_{8}$, excluding large positive values of $x_{6}$. A natural
scaling further points towards $\left|  G_{8}^{s}/G_{8}\right|  <1$, which
would guarantee $x_{i}\sim\mathcal{O}(1)$ and also preserve the sign of
$x_{6}$ with respect to the factorization approximation, i.e., $x_{6}>0$. A
small fraction of penguins, as in the factorization approximation at 1 GeV
where $\mathcal{F}_{P}$ is around $20\%$, would require $x_{6}\simeq1$, i.e.,
a ratio $G_{8}^{s}/G_{8}$ as negative as $-0.8$. However, another theoretical
clue can be obtained from the nonlinear relation between the Wilson
coefficients $z_{1}(\mu)$ and $z_{2}(\mu)$ derived in the leading logarithmic
approximation \cite{WeakHam}:%
\begin{equation}
\left(  z_{2}+z_{1}\right)  ^{2}\left(  z_{2}-z_{1}\right)  =1.
\end{equation}
Indeed, this relation is known to receive rather small corrections at
next-to-leading order (even for $\mu$ as low as $700$ MeV). Assuming%
\begin{equation}
\left(  x_{2}+x_{1}\right)  ^{2}\left(  x_{2}-x_{1}\right)  =\frac{25G_{27}%
^{2}\left(  G_{27}-6G_{8}^{s}\right)  }{27G_{W}^{3}}\simeq1 \label{LeadLog}%
\end{equation}
to within, say, $30\%$, would then give the smaller (in absolute value) ratio:%
\begin{equation}
\left(  G_{8}^{s}/G_{8}\right)  _{th}=-0.38\pm0.12. \label{QCDinspG8s}%
\end{equation}

\subsection{Tracking down penguins at the poles}

Let us now tackle the phenomenological determination of $G_{8}^{s}$. The
vertices induced by $Q_{8}^{s}$ always involve one single $\eta_{0}$ field
(see Eq.(\ref{Q8sStructure})). It is thus natural to turn to $\eta_{0}$ pole
contributions to anomaly-driven radiative kaon decays. These proceed through
the $\mathcal{O}(p^{4})$ Wess-Zumino-Witten (WZW) action of $U(3)$ ChPT, which
is for QED external sources\cite{WZW}:%
\begin{align}
\mathcal{L}_{WZW}^{1\gamma}  &  =\frac{N_{c}}{48\pi^{2}}e\;\varepsilon^{\mu
\nu\rho\sigma}A_{\mu}\langle\partial_{\nu}U\partial_{\rho}U^{\dagger}%
\partial_{\sigma}U\{U^{\dagger},Q\}\rangle\,,\label{WZW1}\\
\mathcal{L}_{WZW}^{2\gamma}  &  =\frac{iN_{c}}{48\pi^{2}}e^{2}\;\varepsilon
^{\mu\nu\rho\sigma}F_{\mu\nu}A_{\rho}\left(  \langle QQ\{\partial_{\sigma
}U,U^{\dagger}\}\rangle+\frac{1}{2}\langle QU^{\dagger}Q\partial_{\sigma
}U-QUQ\partial_{\sigma}U^{\dagger}\rangle\right)  \,, \label{WZW2}%
\end{align}
with $A_{\mu}$, the electromagnetic field, $F_{\mu\nu}\equiv\partial_{\mu
}A_{\nu}-\partial_{\nu}A_{\mu}$ the corresponding strength tensor,
$Q=diag(Q_{u},Q_{d},Q_{s})$, the light quark charge matrix in units of the
positron charge $e$ and $\varepsilon^{\mu\nu\rho\sigma}$, the Levi-Civita
tensor with $\varepsilon^{0123}=+1$. Note that in $U(3)$ ChPT, there are
additional unnatural parity operators at $\mathcal{O}(p^{4})$ (necessarily
chiral invariant as unrelated to the WZW action), like for example
\begin{equation}
\varepsilon^{\mu\nu\rho\sigma}F_{\mu\nu}A_{\rho}\left\langle QQ\right\rangle
\langle\partial_{\sigma}UU^{\dagger}\rangle\,, \label{WZW3}%
\end{equation}
but, being $1/N_{c}$ suppressed, they will be discarded. This is supported by
the analysis of the decays $\eta,\eta^{\prime}\rightarrow\gamma\gamma$,
reasonably well reproduced at leading order by Eqs.(\ref{WZW2}) and
(\ref{LagrNL}) alone\cite{Leutwyler97}.

In the next sections, we will consider pole-dominated radiative modes like
$K_{L}\rightarrow\gamma\gamma$, $K_{S}\rightarrow\pi^{0}\gamma\gamma$ and
$K_{L}\rightarrow\pi^{0}\pi^{0}\gamma\gamma$, as well as transitions that
receive also other types of contributions like $K^{+}\rightarrow\pi^{+}%
\gamma\gamma$ and $K_{L}\rightarrow\pi^{+}\pi^{-}\gamma$.

\section{The $K_{L}\rightarrow\gamma\gamma$ decay}

Despite a long history, and precise measurements by the NA48 and KLOE
Collaborations,%
\begin{equation}
\frac{\Gamma\left(  K_{L}\rightarrow\gamma\gamma\right)  }{\Gamma\left(
K_{L}\rightarrow\pi^{0}\pi^{0}\pi^{0}\right)  }=\left\{
\begin{array}
[c]{l}%
\left(  2.81\pm0.01_{stat}\pm0.02_{syst}\right)  \times10^{-3}%
\text{\cite{NA48KL}}\\
\left(  2.79\pm0.02_{stat}\pm0.02_{syst}\right)  \times10^{-3}%
\text{\cite{KLOEKL}}%
\end{array}
\right.  \;, \label{KL1}%
\end{equation}
it is fair to say that a satisfactory theoretical description of the decay
$K_{L}\rightarrow\gamma\gamma$ is still lacking.

At lowest order in the chiral expansion, i.e. $\mathcal{O}(p^{4})$, it
proceeds through pseudoscalar poles, as depicted in Fig.2a. In the $SU(3)$
framework, the $\eta_{0}$ is a higher order effect and only the $\pi^{0}$ and
$\eta_{8}$ are allowed to propagate, with the result that the amplitude
vanishes exactly:
\begin{equation}
\mathcal{A}^{\mu\nu}\left(  K_{L}\rightarrow\gamma\gamma\right)
\overset{SU(3)}{=}\frac{2F\alpha}{\pi}\left(  G_{27}-G_{8}+G_{8}^{m}\right)
m_{K}^{2}\left(  \frac{1}{m_{K}^{2}-m_{\pi}^{2}}+\frac{1/3}{m_{K}^{2}%
-m_{\eta8}^{2}}\right)  i\varepsilon^{\mu\nu\rho\sigma}k_{1\rho}k_{2\sigma
}=0\;, \label{KL2}%
\end{equation}
upon enforcing the Gell-Mann--Okubo (GMO) mass relation (valid at this order):%
\begin{equation}
m_{\eta8}^{2}=\frac{4m_{K}^{2}-m_{\pi}^{2}}{3}\;. \label{KL3}%
\end{equation}

The $SU(3)$ decay amplitude therefore starts at $\mathcal{O}(p^{6})$, with in
particular all the effects of the $\eta_{0}$. At this stage, phenomenological
pole models are constructed
\cite{DummPich,PhenoPole,MaPramudita81,GaillardL74,Cheng90}, involving the
physical $\eta^{(\prime)}$ masses but also chiral symmetry breaking parameters
for the weak transitions $K_{L}\rightarrow\pi^{0},\eta,\eta^{\prime}$ and the
radiative decays $\pi^{0},\eta,\eta^{\prime}\rightarrow\gamma\gamma$. Yet no
definite prediction can be attained because, having a vanishing lowest order,
the rate exhibits a high sensitivity to these phenomenological parameters.%
\begin{figure}
[t]
\begin{center}
\includegraphics[
height=1.2661in,
width=4.7781in
]%
{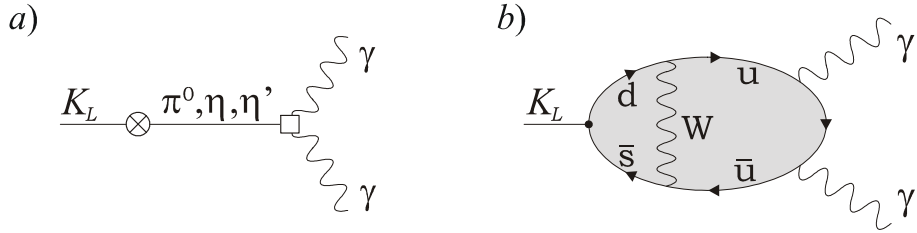}%
\caption{a) Pole diagrams for $K_{L}\rightarrow\gamma\gamma$. b) Dominant
long-distance $\bar{u}u$ contribution.}%
\end{center}
\end{figure}

\subsection{Pole amplitude in U(3) ChPT}

In the present work, large-$N_{c}$ ChPT is used to introduce the effects of
the $\eta^{\prime}$ state consistently in both the weak and strong Lagrangians
Eqs.(\ref{LagrNL}), (\ref{LagrWspurion}) and (\ref{WZW2}). The amplitude is
then non-vanishing already at $\mathcal{O}(p^{4})$, featuring the parameter of
interest $G_{8}^{s}$:%
\begin{equation}
\mathcal{A}^{\mu\nu}\left(  K_{L}\rightarrow\gamma\gamma\right)
\overset{U(3)}{=}\frac{16F\alpha}{\pi}\left(  G_{8}^{s}+\frac{2}{3}%
G_{27}\right)  \left(  \frac{-m_{K}^{2}}{m_{0}^{2}-3m_{K}^{2}+3m_{\pi}^{2}%
}\right)  i\varepsilon^{\mu\nu\rho\sigma}k_{1\rho}k_{2\sigma}\;. \label{KL4}%
\end{equation}
Yet the contributions of the $Q_{8}$ and $Q_{8}^{m}$ operators cancel again
once the $\eta_{8}-\eta_{0}$ theoretical mass matrix is inserted
\begin{equation}
m_{\eta_{8}\eta_{0}}^{2}=\left(
\begin{array}
[c]{cc}%
\dfrac{4m_{K}^{2}-m_{\pi}^{2}}{3} & \dfrac{2\sqrt{2}}{3}(m_{\pi}^{2}-m_{K}%
^{2})\\
\dfrac{2\sqrt{2}}{3}(m_{\pi}^{2}-m_{K}^{2}) & m_{0}^{2}+\dfrac{2m_{K}%
^{2}+m_{\pi}^{2}}{3}%
\end{array}
\right)  \;, \label{KL5}%
\end{equation}
which corresponds to the generalization of GMO to $U(3)$.

Looking back at Eq.(\ref{ChangeBasis}), $G_{8}^{s}+\frac{2}{3}G_{27}$ is
precisely the weight of the $\hat{Q}_{1}$ operator. This is the only one able
to generate a $\bar{u}u$ pair from the incoming $K_{L}$ at lowest order
($\hat{Q}_{3,5}$ are again expected to give small contributions, and are
discarded). The flavor and color structures of $Q_{2}$ prevent $\hat{Q}_{2}$
from contributing, while $\hat{Q}_{6}$ appears only through its $\bar{d}%
d+\bar{s}s$ component and cancels out\footnote{Such a cancellation of the
$Q_{6}$ penguin operator has already been noticed in the $m_{0}\rightarrow0$
limit \cite{BurasGerard86}.}. The decay $K_{L}\rightarrow\gamma\gamma$ thus
proceeds entirely through the long-distance $\bar{s}d(\overline{d}%
s)\rightarrow\bar{u}u\rightarrow\gamma\gamma$ transition at lowest order
(Fig.2b), which is in agreement with a quark-level
analysis\cite{MaPramudita81}.

Let us stress once more why working in $U(3)$ is essential. Beside the fact
that the quark basis $\bar{u}u,\bar{d}d,\bar{s}s$ is ambiguous in $SU(3)$,
there are not enough $|\Delta S|=1$ operators to permit a definite
identification of the underlying transitions in this symmetry limit. In fact,
$\hat{Q}_{1},\hat{Q}_{2}$ and $\hat{Q}_{6}$ are not independent in $SU(3)$,
hence the total $K_{L}\rightarrow\gamma\gamma$ amplitude has to vanish at
$\mathcal{O}(p^{4})$ since the last two do not contribute.

The above interpretation can be made obvious by a direct computation in the
quark basis:%
\begin{equation}
\mathcal{A}^{\mu\nu}\left(  K_{L}\rightarrow\gamma\gamma\right)
\overset{U(3)}{=}V_{weak}^{T}\cdot P_{LO}\left(  m_{K}^{2}\right)  \cdot
V_{\gamma\gamma}^{\mu\nu}\,, \label{KL6}%
\end{equation}
with the LO propagator ($R\equiv2N_{c}(m_{K}^{2}-m_{\pi}^{2})/m_{0}^{2}\,$):%
\begin{equation}
iP_{LO}\left(  q^{2}\right)  ^{-1}=\left(  q^{2}-m_{\pi}^{2}\right)  \left(
\begin{array}
[c]{ccc}%
1 & 0 & 0\\
0 & 1 & 0\\
0 & 0 & 1
\end{array}
\right)  -\frac{m_{0}^{2}}{N_{c}}\left(
\begin{array}
[c]{ccc}%
1 & \;\;1 & 1\\
1 & \;\;1 & 1\\
1 & \;\;1 & 1+R
\end{array}
\right)  \,, \label{KL7}%
\end{equation}
the WZW $\gamma\gamma$ vertices (proportional to $Q_{q}^{2}$):%
\begin{equation}
V_{\gamma\gamma}^{\mu\nu}=-\frac{\sqrt{2}N_{c}\alpha}{\pi F}\left(
\begin{array}
[c]{c}%
4/9\\
1/9\\
1/9
\end{array}
\right)  i\varepsilon^{\mu\nu\rho\sigma}k_{1\rho}k_{2\sigma}\,, \label{KL8}%
\end{equation}
and the weak vertices, in both the $(Q_{8},Q_{8}^{s},Q_{27})$ and $(\hat
{Q}_{1},\hat{Q}_{2},\hat{Q}_{6})$ bases (still keeping $Q_{8}^{m}$):%
\begin{equation}
V_{weak}=i2\sqrt{2}F^{2}m_{K}^{2}\left(
\begin{tabular}
[c]{r}%
$G_{8}^{s}+2G_{27}/3$\\
$G_{8}-G_{8}^{m}+G_{8}^{s}-G_{27}/3\;\,$\\
$G_{8}-G_{8}^{m}+G_{8}^{s}-G_{27}/3\;\,$%
\end{tabular}
\right)  =i2\sqrt{2}F^{2}m_{K}^{2}G_{W}\left(
\begin{tabular}
[c]{r}%
$x_{1}$\\
$x_{6}$\\
$x_{6}$%
\end{tabular}
\right)  \;. \label{KL9}%
\end{equation}
The $\bar{d}d$ and $\bar{s}s$ components of $V_{weak}$ are equal due to $CPS$
invariance. One can easily check that only the $\bar{u}u$ component of the
weak vector contributes to the amplitude, as announced.

A priori, one could think that NLO effects may reintroduce significant
$\hat{Q}_{6}$ contributions. This intuition is however based on the
pathological $SU(3)$ situation, in which there is an extreme sensitivity due
to the inability of $SU(3)$ to catch the leading order $\bar{s}d(\overline
{d}s)\rightarrow\bar{u}u\rightarrow\gamma\gamma$ contribution. Once within
$U(3)$, this piece is correctly identified and we can consider that NLO
effects will behave according to the usual chiral counting, i.e. are at most
of $30\%$. In other words, we will work under the assumption that $\hat{Q}%
_{1}$ represents the dominant contribution to the decay.

\subsection{Physical mass prescription}

Diagonalizing Eq.(\ref{KL5}), the leading order $\eta$ and $\eta^{\prime}$
theoretical (or Lagrangian) masses are found to be%
\begin{equation}
m_{\eta(\eta^{\prime})}^{2}=\frac{m_{0}^{2}}{2}+m_{K}^{2}\pm\frac{1}{2}%
\sqrt{m_{0}^{4}-\frac{4}{3}(m_{K}^{2}-m_{\pi}^{2})m_{0}^{2}+4(m_{K}^{2}%
-m_{\pi}^{2})^{2}}\;. \label{PMP0}%
\end{equation}
The parameter $m_{0}$ has to be fitted to reproduce both $m_{\eta}$ and
$m_{\eta^{\prime}}$ as close as possible to their physical values, which gives
$m_{0}$ between $800$ and $900$ MeV. From Fig.3, the masses are then well
reproduced, within less than $15\%$, which is quite consistent with chiral
counting for higher order effects.%

\begin{figure}
[t]
\begin{center}
\includegraphics[
height=1.6267in,
width=6.2509in
]%
{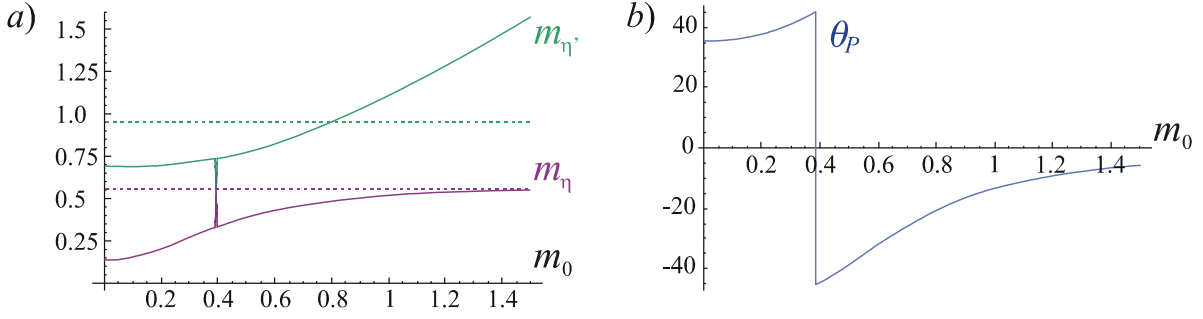}%
\caption{$\eta-\eta^{\prime}$ masses (in GeV) and mixing angle at lowest order
in large-$N_{c}$ $U(3)$ ChPT.}%
\end{center}
\end{figure}

Still, in the present work, we are mainly concerned by pole amplitudes, i.e.
amplitudes involving virtual $\eta-\eta^{\prime}$ exchanges. For such
processes, a good physical mass prescription must ensure correct analytical
properties, i.e., freeze the poles at their right places. Varying $m_{0}$ is
therefore not satisfactory. In analogy with the usual $\mathcal{O}(p^{2})$
substitutions
\begin{align}
m_{\pi}^{2}  &  =\frac{r}{2}\left(  m_{u}+m_{d}\right)  \rightarrow M_{\pi
}^{2}\simeq(135\text{ MeV})^{2}\text{\thinspace,}\nonumber\\
m_{K}^{2}  &  =\frac{r}{2}\left(  m_{s}+m_{u,d}\right)  \rightarrow M_{K}%
^{2}\simeq(495\text{ MeV})^{2}\text{\thinspace,} \label{PMP1}%
\end{align}
our prescription is to substitute for the LO mass matrix Eq.(\ref{KL5}) the
most general mass matrix compatible with isospin symmetry\footnote{In
principle, the pseudoscalar widths have to be included to really ensure
correct analytical properties. However, these are numerically not relevant and
will be neglected.}. In the $\bar{u}u,\bar{d}d,\bar{s}s$ basis considered in
Eq.(\ref{KL7}), the ($\pi^{0},\eta_{8},\eta_{0}$) propagator takes then the
generic form
\begin{equation}
iP_{phys}\left(  q^{2}\right)  _{\bar{q}q}^{-1}=\left(  q^{2}-M_{\pi}%
^{2}\right)  \left(
\begin{array}
[c]{ccc}%
1 & 0 & 0\\
0 & 1 & 0\\
0 & 0 & 1
\end{array}
\right)  -\frac{M_{0}^{2}}{3}\left(
\begin{array}
[c]{ccc}%
1 & 1 & 1-\delta\\
1 & 1 & 1-\delta\\
1-\delta & 1-\delta & 1+\bar{R}-2\delta
\end{array}
\right)  \,, \label{PMP2}%
\end{equation}
with $M_{0},\delta,\bar{R},$ some parameters to be adjusted to freeze the
poles at their physical values. Equivalently, working in the $\eta_{8}%
,\eta_{0}$ basis, this propagator can be written%
\begin{equation}
iP_{phys}\left(  q^{2}\right)  _{\eta_{8}\eta_{0}}^{-1}=\left(
\begin{array}
[c]{cc}%
\cos\theta_{P} & \sin\theta_{P}\\
-\sin\theta_{P} & \cos\theta_{P}%
\end{array}
\right)  \left(
\begin{array}
[c]{cc}%
q^{2}-M_{\eta}^{2} & 0\\
0 & q^{2}-M_{\eta^{\prime}}^{2}%
\end{array}
\right)  \left(
\begin{array}
[c]{cc}%
\cos\theta_{P} & -\sin\theta_{P}\\
\sin\theta_{P} & \cos\theta_{P}%
\end{array}
\right)  \,, \label{PMP3}%
\end{equation}
and the $\pi^{0}$ does not mix (for the model-independent relationship between
Eqs.(\ref{PMP2}) and (\ref{PMP3}), see Ref.\cite{GerardK05}). The masses are
then fixed to%
\begin{equation}
M_{\eta}\simeq547.8\text{ MeV,\quad}M_{\eta^{\prime}}\simeq957.8\text{ MeV,}
\label{PMP3b}%
\end{equation}
with the mixing angle around $\theta_{P}\simeq-22%
{{}^\circ}%
$ if a large-$N_{c}$ limit is taken at each order in the momentum expansion
\cite{GerardK05}.

When the leading order is non-zero, inserting physical masses amounts to
including a particular class of higher order effects. Yet there are still many
other sources of corrections, and to get a handle on their order of magnitude,
we will consider $\theta_{P}$ as an effective angle, allowed to vary around
$-22%
{{}^\circ}%
$. Specifically, the sensitivity of the pole amplitudes will be probed by
taking $\theta_{P}$ inside the range [$-15%
{{}^\circ}%
,-25%
{{}^\circ}%
$].

For example, we could include, in addition to the mass corrections, the
deviations from $F_{\eta}=F_{\eta^{\prime}}=F_{\pi}$ by introducing the
renormalized decay constant matrix $\mathbf{F}_{\eta}$ as%
\begin{equation}
P_{phys}\left(  q^{2}\right)  _{\eta_{8}\eta_{0}}^{\prime}=F^{2}\;\left(
\mathbf{F}_{\eta}\right)  ^{-1}.\left(  P_{phys}\left(  q^{2}\right)
_{\eta_{8}\eta_{0}}\right)  .\left(  \mathbf{F}_{\eta}^{T}\right)  ^{-1}\,.
\label{PMP4}%
\end{equation}
This amounts to switching to a two-mixing angle formalism since by definition
\begin{equation}
\left(
\begin{array}
[c]{cc}%
\cos\theta_{P} & -\sin\theta_{P}\\
\sin\theta_{P} & \cos\theta_{P}%
\end{array}
\right)  \mathbf{F}_{\eta}\equiv\left(
\begin{array}
[c]{cc}%
F_{8}^{\eta} & F_{0}^{\eta}\\
F_{8}^{\eta^{\prime}} & F_{0}^{\eta^{\prime}}%
\end{array}
\right)  \equiv\left(
\begin{array}
[c]{cc}%
F_{8}\cos\theta_{8} & -F_{0}\sin\theta_{0}\\
F_{8}\sin\theta_{8} & F_{0}\cos\theta_{0}%
\end{array}
\right)  \;. \label{PMP5}%
\end{equation}
However, not much is gained by using this prescription instead of
Eq.(\ref{PMP3}) as there are still unaccounted NLO effects beyond masses and
decay constants. Numerically, we have checked that in all the cases
considered, taking Eq.(\ref{PMP5}) with the values of $F_{8}$, $F_{0}$,
$\theta_{8}$ and $\theta_{0}$ from either the large $N_{c}$ analysis of
Ref.\cite{Leutwyler97} or the phenomenological fit of Ref.\cite{EscribanoF05}
does not alter our conclusions (in general, the deviation with respect to
Eq.(\ref{PMP3}) with $\theta_{P}\simeq-22%
{{}^\circ}%
$ can be accounted for by taking $\theta_{P}\simeq-15%
{{}^\circ}%
$ or slightly smaller).

\subsection{The $K_{L}\rightarrow\gamma\gamma$ rate and its implications for
the $\Delta I=1/2$ rule}

Naively enforcing the physical mass prescription, i.e. taking Eq.(\ref{KL6})
with the propagator Eq.(\ref{PMP2}), would however reintroduce a large
$\hat{Q}_{6}$ contribution. This is a spurious effect because it corresponds
to the partial inclusion of higher order terms (only those from mass
corrections) for a contribution which vanishes at lowest order. As said
before, the decay should be dominated by the $\bar{u}u$ transition, hence
large cancellations are expected to occur for $\hat{Q}_{6}$ at NLO
(indications of these are discussed in the next subsection).

On the contrary, since the $\bar{u}u$ part is non-vanishing at lowest order,
inserting the physical masses instead of the Lagrangian ones is allowed.
Therefore, we definitively discard the $\bar{d}d$ and $\bar{s}s$ components
from the weak vertex:
\begin{equation}
V_{weak}=i2\sqrt{2}F^{2}m_{K}^{2}G_{W}\left(
\begin{array}
[c]{c}%
x_{1}\\
0\\
0
\end{array}
\right)  =i2\sqrt{2}F^{2}m_{K}^{2}\left(
\begin{array}
[c]{c}%
G_{8}^{s}+\dfrac{2}{3}G_{27}\\
0\\
0
\end{array}
\right)  \,, \label{KL11}%
\end{equation}
and use Eq.(\ref{PMP2}) to reach the final form for the amplitude ($c_{\theta
}\equiv\cos\theta_{P},s_{\theta}\equiv\sin\theta_{P}$):%
\begin{align}
&  \mathcal{A}^{\mu\nu}\left(  K_{L}\rightarrow\gamma\gamma\right)
=\frac{2F\alpha}{\pi}\left(  G_{8}^{s}+\dfrac{2}{3}G_{27}\right)  M_{K}%
^{2}i\varepsilon^{\mu\nu\rho\sigma}k_{1\rho}k_{2\sigma}\nonumber\\
&  \;\;\;\times\left(  \frac{1}{M_{K}^{2}-M_{\pi}^{2}}+\frac{(c_{\theta
}-2\sqrt{2}s_{\theta})(c_{\theta}-\sqrt{2}s_{\theta})}{3(M_{K}^{2}-M_{\eta
}^{2})}+\frac{(s_{\theta}+2\sqrt{2}c_{\theta})(s_{\theta}+\sqrt{2}c_{\theta}%
)}{3(M_{K}^{2}-M_{\eta^{\prime}}^{2})}\right)  \;. \label{KL12}%
\end{align}

Numerically, this amplitude is dominated by the $\eta$ pole:%
\begin{equation}
\mathcal{A}^{\mu\nu}\left(  K_{L}\rightarrow\gamma\gamma\right)  =\left(
G_{8}^{s}+\dfrac{2}{3}G_{27}\right)  \left[  \left(  0.46\right)  _{\pi
}-\left(  1.83\pm0.30\right)  _{\eta}-\left(  0.12\pm0.02\right)
_{\eta^{\prime}}\right]  i\varepsilon^{\mu\nu\rho\sigma}k_{1\rho}k_{2\sigma
}\;, \label{KL13}%
\end{equation}
where the numbers are in MeV and the errors amount to varying $\theta_{P}$
between $-15%
{{}^\circ}%
$ and $-25%
{{}^\circ}%
$. The parametrization is rather stable with respect to $\eta-\eta^{\prime}$
mixing parameters, which means that the amplitude is under control once the
poles are frozen at their right places. Then, from the experimental values
Eqs.(\ref{KL1}) with $\Gamma(K_{L}\rightarrow\pi^{0}\pi^{0}\pi^{0})$ from
\cite{KLOE,PDG04}, the only free parameter can be extracted up to a two-fold
ambiguity:
\begin{equation}
\left(  G_{8}^{s}/G_{8}\right)  _{ph}\simeq\pm1/3\;. \label{KL14}%
\end{equation}%

\begin{table}[t] \centering
\begin{tabular}
[c]{c|c|cccc|cc}\hline
$\theta_{P}(%
{{}^\circ}%
)$ & $G_{8}^{s}/G_{8}$ & $x_{1}$ & $x_{2}$ & $x_{6}$ & $\left(  x_{1}%
+x_{2}\right)  ^{2}\left(  x_{2}-x_{1}\right)  $ & $\mathcal{F}_{CC}(\%)$ &
$\mathcal{F}_{P}(\%)$\\\hline
$-10$ & $-0.44$ & $-2.06$ & $2.56$ & $2.78$ & $1.15$ & $46$ & $54$\\
$-15$ & $-0.35$ & $-1.59$ & $2.08$ & $3.26$ & $0.91$ & $37$ & $63$\\
$-17.5$ & $-0.32$ & $-1.42$ & $1.92$ & $3.42$ & $0.83$ & $34$ & $66$\\
$-20$ & $-0.29$ & $-1.29$ & $\;1.78\;$ & $\;3.56\;$ & $0.76$ & $31$ & $69$\\
$-22.5$ & $-0.27$ & $-1.18$ & $1.67$ & $3.67$ & $0.71$ & $29$ & $71$\\
$-25$ & $-0.25$ & $-1.08$ & $1.58$ & $3.76$ & $0.66$ & $27$ & $73$\\
$-30$ & $-0.22$ & $-0.94$ & $1.44$ & $3.90$ & $0.59$ & $25$ & $75$\\\hline
\end{tabular}
\caption{The coupling $G_8^s$ as extracted from $K_{L}%
\rightarrow\gamma
\gamma$, in terms of $\theta_{P}%
$, and the corresponding weights $x_{i}%
$ for the 
current $\times$ current and penguin operators. The last two columns give the 
current $\times$ current and penguin fractions in $A_0$ defined in Eq.(\ref{CCPFrac}%
).
\label{Table1}}
\end{table}%

From QCD-inspired arguments, the positive solution is discarded as it would
lead to $x_{1},$ $x_{2}$ of the wrong signs and $x_{6}$ very large (see
Fig.1). The negative solution%
\begin{equation}
\left(  G_{8}^{s}/G_{8}\right)  _{ph+th}=-0.30\pm0.05 \label{G8sPhTh}%
\end{equation}
is detailed in Table \ref{Table1} and Fig.4. Also, $\left(  x_{1}%
+x_{2}\right)  ^{2}\left(  x_{2}-x_{1}\right)  $ stays remarkably close to
one, which means that the low energy evolution of current $\times$ current
operators is indeed rather smooth.

On the other hand, $x_{6}$ increases quite dramatically from its perturbative
QCD value Eq.(\ref{EqNorm}) and appears responsible for about $2/3$ of the
$\Delta I=1/2$ enhancement%
\begin{equation}
\mathcal{F}_{P}=\left(  68\pm6\right)  \% \label{Peng}%
\end{equation}
where the error reflects only the variation of $\theta_{P}$. Let us stress
again that $x_{1}$, $x_{2}$ and $x_{6}$ are the weights of the weak operators
at a very low hadronic scale, around the ChPT scale $F$. It is intuitively
clear then that $x_{6}$, and thus $\mathcal{F}_{P}$, can be large since we
allow the evolution $Q_{1},Q_{2}\rightarrow Q_{6}$ all the way down to this scale.%

\begin{figure}
[t]
\begin{center}
\includegraphics[
height=1.7149in,
width=6.2604in
]%
{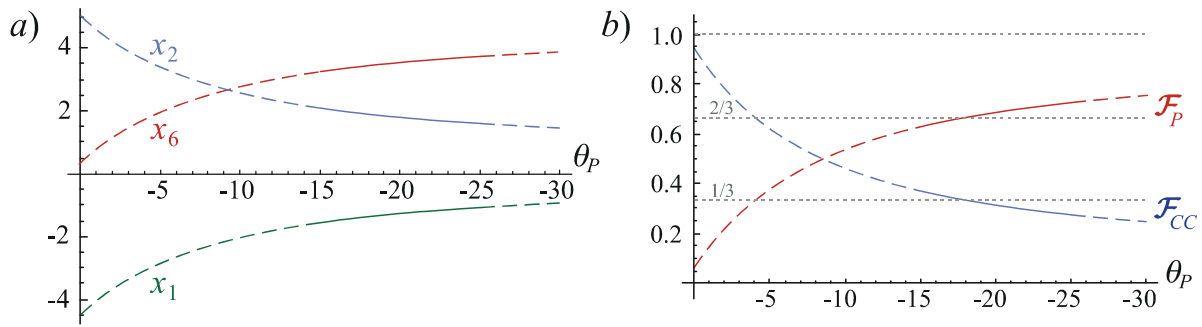}%
\caption{Graphical representation of Table 1: a) Weak operator weights $x_{i}$
as a function of $\theta_{P}$, b) Corresponding current $\times$ current and
penguin fractions in $A_{0}$.}%
\end{center}
\end{figure}

\subsection{Effective suppression of $Q_{6}$ at higher order}

In the context of $SU(3)$, there is only one weak octet operator, hence it is
clear that large $\mathcal{O}(p^{6})$ \textquotedblleft$Q_{8}$%
\textquotedblright\ effects are present. Indeed, it is through the
counterterms that the leading $\bar{s}d(\overline{d}s)\rightarrow
\bar{u}u\rightarrow\gamma\gamma$ transition is reconstructed. Yet, this
reconstruction is not trivial and implies a specific interplay between strong
and weak counterterms of both $\mathcal{O}(p^{4})$ and $\mathcal{O}(p^{6})$,
corresponding to corrections to the decay constants, two-photon vertices,
masses (i.e., corrections to GMO) and weak transitions.

In the appendix, the complete analysis of $K_{L}\rightarrow\gamma\gamma$ at
$\mathcal{O}(p^{6})$ in $SU(3)$ ChPT is presented. It is shown that, to a
large extent, the counterterms either act together to reconstruct $\hat{Q}%
_{1}$, or cancel among themselves. For example, corrections brought by
$F_{\eta8}\neq F_{\pi}$ cancel out with some mass corrections, and are thus at
least of $\mathcal{O}(p^{8})$. Only a small irreducible combination of
(scalar-saturated) counterterms survives, on which not much can be said at
present. Still, given the dynamical insight gained, it is reasonable to expect
this combination to be small. In conclusion, there is a clear indication that
no large contribution from $\hat{Q}_{6}$ is generated at $\mathcal{O}(p^{6})$
in $SU(3)$ ChPT.

Something very similar should happen in $U(3)$, but it is unfortunately
difficult to be as quantitative since we do not have any handle on the $U(3)$
weak counterterms. Anyway, here also, it is likely that corrections to the
masses, decay constants, weak and WZW vertices cancel among themselves, at
least in part, for the $\hat{Q}_{6}$ contribution.

This is to be contrasted to the phenomenological pole model
analyses\cite{DummPich,PhenoPole}, which overlook cancellations among
$\mathcal{O}(p^{6})$ corrections. To be more specific, these models usually
introduce a parameter $\rho$ to account for nonet symmetry breaking. In
$U(3)$, this breaking is entirely due to the $Q_{8}^{s}$ and $Q_{27}$
operators (which induce only $K_{L}\rightarrow\eta_{0}$ and $K_{L}%
\rightarrow\pi^{0},\eta_{8}$ transitions, respectively):%
\begin{equation}
\rho-1\equiv-\frac{1}{2}\sqrt{\frac{3}{2}}\frac{\left\langle \eta_{0}\left|
\mathcal{H}_{W}^{\Delta S=1}\right|  K_{L}\right\rangle }{\left\langle \pi
^{0}\left|  \mathcal{H}_{W}^{\Delta S=1}\right|  K_{L}\right\rangle
}-1=\frac{3}{2}\frac{G_{8}^{s}+2G_{27}/3}{G_{8}-G_{27}}=\frac{3}{2}%
\frac{x_{1}}{x_{6}-x_{1}}\ . \label{KL15}%
\end{equation}
Obviously, the nonet symmetry limit, $\rho=1$, amounts to discarding the
dominant $\hat{Q}_{1}$ contribution (i.e. $K_{L}\rightarrow\bar{u}u\rightarrow
\gamma\gamma$), and keeping only the $\hat{Q}_{6}$ one. Therefore, in that
limit, these models should under-estimate the $K_{L}\rightarrow\gamma\gamma$
rate, but this is not the case (on the contrary, they need $\rho<1$ to
\textit{reduce} the $\hat{Q}_{6}$ contribution). In other words, from the
dominance of the $K_{L}\rightarrow\bar{u}u\rightarrow\gamma\gamma$ transition,
a correct phenomenological pole model should be proportional to $\rho-1$ (like
Eq.(\ref{KL12})), to a good approximation.

\subsection{Comments on $K_{L}\rightarrow\gamma\gamma^{\ast}$}

For the $K_{L}\rightarrow\gamma\ell^{+}\ell^{-}$ decays, a detailed slope
analysis was performed in Ref.\cite{DAmbrosioP97KLgg}, including vector meson
exchanges. Two remarks can be made from the insight gained here.

First, quite generally, our two-step phenomenological procedure to deal with
pole amplitudes in $U(3)$ could have some implications also for VMD-type
models. Indeed, working first with theoretical masses permits the
identification of vanishing contributions, for which it is very dangerous to
include only partially the higher order corrections (due only to the masses
for example). As a rule, cancellations plague pole amplitudes and should be
dealt with carefully. This is especially true in $SU(3)$ ChPT, since it is
unable to catch a transition through a $\bar{u}u$ quark pair at leading order.
Only once the surviving terms are precisely identified, one is allowed to
switch to physical masses without risks.

Given the specific topologies studied in Ref.\cite{DAmbrosioP97KLgg}, it is
the parameter $b_{V}^{nonet}$ that could appear most sensitive to the previous
remark. It originates from the processes $K_{L}\rightarrow\pi^{0},\eta
,\eta^{\prime}\rightarrow\gamma(\rho,\omega,\phi)\rightarrow\gamma\gamma
^{\ast}$, which vanish at lowest order in $SU(3)$ ChPT, and proceed through
$G_{8}^{s}+\frac{2}{3}G_{27}$ in $U(3)$ ChPT. However, all the dependences on
the weak transition cancel in the normalization with respect to $K_{L}%
\rightarrow\gamma\gamma$, and $b_{V}^{nonet}$ as given in
Ref.\cite{DAmbrosioP97KLgg} is not modified.

Second, it is to be noted that the global sign of our $K_{L}\rightarrow
\gamma\gamma$ amplitude when $G_{8}^{s}/G_{8}<0$ is opposite to the one
obtained using phenomenological pole models\footnote{These models in general
imply a destructive interference between the $\eta$ and $\eta^{\prime}$ poles,
leaving the $\pi^{0}$ one as dominant (see the discussion in \cite{DummPich}).
Then, one can see (independently of our conventions) that the sign of the
$\pi^{0}$ pole in Eq.(\ref{KL2}), keeping only $G_{8}$, is the opposite of the
$\eta$ one of Eq.(\ref{KL13}) when $G_{8}^{s}/G_{8}<0$.}. This is supported by
the analysis of Ref.\cite{DAmbrosioP97KLgg}. Yet, a precise constraint on
$G_{8}^{s}/G_{8}$ from $K_{L}\rightarrow\gamma\gamma^{\ast}$ seems at present
difficult to obtain given the many unknown counterterms and VMD couplings. So,
the purpose of the remaining sections is to search for independent constraints
on $G_{8}^{s}/G_{8}$ from other radiative $K$ decays.

\section{The $K_{S}\rightarrow\pi^{0}\gamma\gamma$ decay}

The radiative process $K_{S}\rightarrow\pi^{0}\gamma\gamma$ is the simplest
mode one can think of to test the physical mechanism advocated for
$K_{L}\rightarrow\gamma\gamma$, with in particular $G_{8}^{s}/G_{8}\simeq
-1/3$. The corresponding branching fraction has been measured two years ago by
the NA48 collaboration, though with large uncertainties \cite{NA48KS}:%
\begin{equation}
\mathcal{B}\left(  K_{S}\rightarrow\pi^{0}\gamma\gamma\right)  _{m_{\gamma
\gamma}>220\text{ MeV}}^{\exp}=\left(  4.9\pm1.8\right)  \times10^{-8}\ .
\label{Ks1}%
\end{equation}%
\begin{figure}
[t]
\begin{center}
\includegraphics[
height=1.2367in,
width=3.8432in
]%
{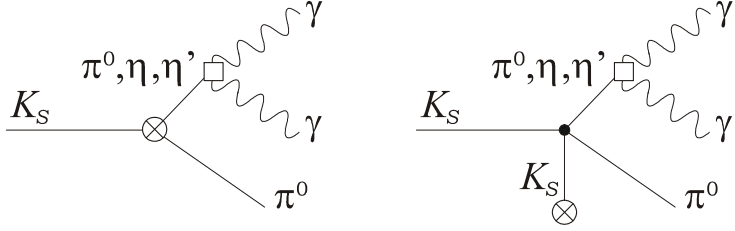}%
\caption{Pole and tadpole diagrams for the process $K_{S}\rightarrow\pi
^{0}\gamma\gamma$.}%
\end{center}
\end{figure}
The decay proceeds through pole diagrams at leading order \cite{EckerPR87},
with also a tadpole graph in the case of the weak mass operator $Q_{8}^{m}$
\cite{BijnensPP98} (Fig.5). Unlike for $K_{L}\rightarrow\gamma\gamma$, the
amplitude starts at $\mathcal{O}(p^{4})$ in both $U(3)$ and $SU(3)$ ChPT. It
can be parametrized in terms of a pole function $B(T^{2})$:%
\begin{equation}
\mathcal{A}^{\mu\nu}\left(  K_{S}\rightarrow\pi^{0}\gamma\gamma\right)
=\frac{8\alpha}{3\pi}B(T^{2})\varepsilon^{\mu\nu\rho\sigma}k_{1\rho}%
k_{2\sigma}\;, \label{Ks2}%
\end{equation}
with $T=k_{1}+k_{2}$, the total momentum of the photon pair. This gives for
the width:%
\begin{equation}
\Gamma\left(  K_{S}\rightarrow\pi^{0}\gamma\gamma\right)  =\frac{2\alpha
^{2}M_{K}^{5}}{9\left(  2\pi\right)  ^{5}}\int_{0.2}^{z_{\max}}dz\lambda
^{1/2}\left(  1,z,r_{\pi}^{2}\right)  z^{2}\left|  B(z)\right|  ^{2}\;,
\label{Ks3}%
\end{equation}
with $z=T^{2}/M_{K}^{2}$, $\lambda\left(  a,b,c\right)  =a^{2}+b^{2}%
+c^{2}-2ab-2ac-2bc$, $r_{\pi}=M_{\pi}/M_{K}$ and $z_{\max}=(1-r_{\pi}%
)^{2}\simeq0.53$. The infrared cut-off is introduced to get rid of the
$K_{S}\rightarrow\pi^{0}\pi^{0}$ background as in Eq.(\ref{Ks1}).

\subsection{Pole amplitude in U(3) ChPT}

A tree-level $U(3)$ computation gives:%
\begin{equation}
B(T^{2})=(m_{K}^{2}-m_{\pi}^{2})\left[  G_{8}B_{8}(T^{2})+G_{8}^{s}B_{8}%
^{s}(T^{2})+G_{27}B_{27}(T^{2})\right]  \label{Ks4}%
\end{equation}
with no contribution from $Q_{8}^{m}$ and%
\begin{gather}
B_{8}(T^{2})=\frac{(m_{K}^{2}-T^{2})(m_{0}^{2}-m_{\pi}^{2}+T^{2})}%
{(T^{2}-m_{\eta}^{2})(T^{2}-m_{\eta^{\prime}}^{2})\left(  T^{2}-m_{\pi}%
^{2}\right)  }\;,\;B_{8}^{s}(T^{2})=\frac{5m_{K}^{2}-2m_{\pi}^{2}-3T^{2}%
}{2(T^{2}-m_{\eta}^{2})(T^{2}-m_{\eta^{\prime}}^{2})}\;,\label{Ks5}\\
B_{27}(T^{2})=\frac{3m_{0}^{2}(T^{2}-m_{K}^{2})+2(m_{K}^{2}-m_{\pi}^{2}%
)(T^{2}-m_{\pi}^{2})}{3(T^{2}-m_{\eta}^{2})(T^{2}-m_{\eta^{\prime}}%
^{2})\left(  T^{2}-m_{\pi}^{2}\right)  }\;.\nonumber
\end{gather}
Note that $m_{\pi,K,\eta,\eta^{\prime}}$ refer to the \textit{theoretical}
masses, defined in Eqs.(\ref{PMP0}) and (\ref{PMP1}). It is only when these
masses are used (i.e., when one works consistently at lowest order) that the
weak mass term $Q_{8}^{m}$ indeed drops out. Note also the identity%
\begin{equation}
\frac{3}{5}B_{8}-\frac{2}{5}B_{8}^{s}+\frac{3}{5}B_{27}=0\;, \label{Ks7}%
\end{equation}
originating from the fact that $\hat{Q}_{2}$ cannot contribute to any of the
subprocesses $K_{S}\rightarrow\pi^{0}(\pi^{0},\eta,\eta^{\prime})$.%
\begin{figure}
[t]
\begin{center}
\includegraphics[
height=1.8049in,
width=6.3071in
]%
{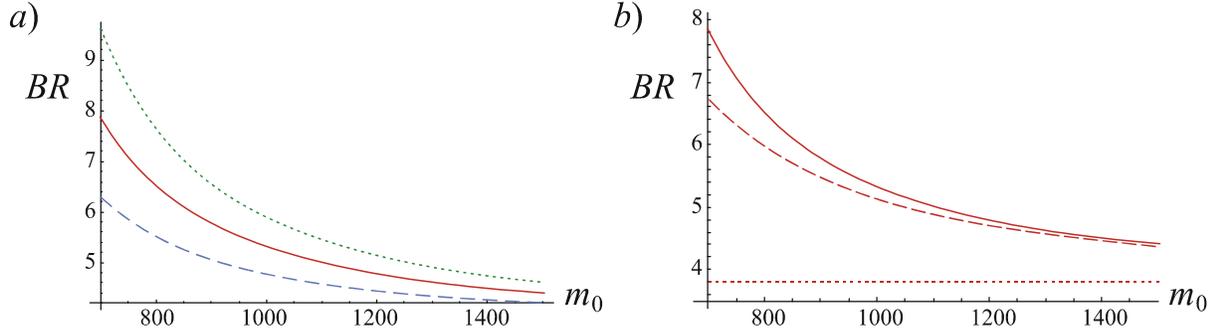}%
\caption{a) $\mathcal{B}\left(  K_{S}\rightarrow\pi^{0}\gamma\gamma\right)  $,
$\times$ $10^{8}$, as a function of $m_{0}$ for $G_{8}^{s}/G_{8}=-0.3,0,+0.3$
(dashed, plain, dotted). b) Comparison between the $\mathcal{O}(p^{4})$
$SU(3)$ result (dotted), idem plus the $m_{0}^{-2}$ corrections (dashed), and
full $\mathcal{O}(p^{4})$ $U(3)$ result (plain), with $G_{8}^{s}=0$ for the
three curves.}%
\end{center}
\end{figure}

The behavior of the rate as a function of $m_{0}$ is plotted in Fig.6a. A
significant enhancement is observed when $m_{0}$ decreases. This is
essentially an effect of the $\eta$ pole, that comes closer to the upper
boundary of phase-space when $m_{0}$ diminishes (Fig.3a). In fact, this pole
could even enter the phase-space if $m_{0}$ was chosen as low as $m_{0}<447$
MeV, which corresponds to $m_{\eta}<m_{K}-m_{\pi}$ from Eq.(\ref{PMP0}). Such
a value is of course to be avoided. The estimation of the actual enhancement
requires the phenomenological prescription proposed in Sec.3.2, and is
deferred to Sec.4.3.

On the other hand, it is clear that, $Q_{8}$ being allowed to contribute at
leading order, it will dominate the decay via the pion pole (and, to a lesser
extent, the $\eta$ one). This explains the moderate sensitivity of
$\mathcal{B}\left(  K_{S}\rightarrow\pi^{0}\gamma\gamma\right)  $ to
$G_{8}^{s}$ exhibited in Fig.6a. However, the decay $K_{S}\rightarrow\pi
^{0}\gamma\gamma$ could still be a useful probe of $G_{8}^{s}/G_{8}$, as will
be discussed in Sec.4.3.

\subsection{Reduction to SU(3) ChPT}

How can the predicted enhancement for decreasing $m_{0}$ (see Fig.6a) be
understood from the point of view of the $SU(3)$ chiral expansion? In order to
answer that question, let us develop the three pole functions (\ref{Ks5}) in
powers of $1/m_{0}^{2}$:%
\begin{equation}%
\begin{array}
[c]{cclc}%
B_{8}\left(  T^{2}\right)  = & \dfrac{T^{2}-m_{K}^{2}}{(T^{2}-m_{\eta8}%
^{2})\left(  T^{2}-m_{\pi}^{2}\right)  } & -\dfrac{2(5m_{K}^{2}-2m_{\pi}%
^{2}-3T^{2})(T^{2}-m_{K}^{2})}{3(T^{2}-m_{\eta8}^{2})^{2}m_{0}^{2}} &
+\mathcal{O}(m_{0}^{-4})\;,\\
B_{8}^{s}\left(  T^{2}\right)  = & 0 & -\dfrac{5m_{K}^{2}-2m_{\pi}^{2}-3T^{2}%
}{2(T^{2}-m_{\eta8}^{2})m_{0}^{2}} & +\mathcal{O}(m_{0}^{-4})\;,\\
B_{27}\left(  T^{2}\right)  = & -\dfrac{T^{2}-m_{K}^{2}}{(T^{2}-m_{\eta8}%
^{2})\left(  T^{2}-m_{\pi}^{2}\right)  } & -\dfrac{(5m_{K}^{2}-2m_{\pi}%
^{2}-3T^{2})(2m_{K}^{2}+m_{\pi}^{2}-3T^{2})}{9(T^{2}-m_{\eta8}^{2})^{2}%
m_{0}^{2}} & +\mathcal{O}(m_{0}^{-4})\;,
\end{array}
\label{Ks8}%
\end{equation}
with $m_{\eta8}^{2}$ fixed by the GMO relation. Though Eq.(\ref{Ks5}) is just
the $\mathcal{O}(p^{4})$ $U(3)$ amplitude, the various terms in the above
series correspond to increasing $p^{2}$ orders in $SU(3)$. As can be checked
numerically, these series are quite well-behaved (i.e., as expected from naive
chiral counting). Still, significant effects can build up through phase-space
integration, leading to the large enhancement observed in Fig.6a. The behavior
of the total rate as a function of $m_{0}$ when the $m_{0}$ series is
truncated at a given order is displayed in Fig.6b:

\begin{quote}
- Retaining only the $\mathcal{O}(m_{0}^{0})$ terms, the $\mathcal{O}(p^{4})$
$SU(3)$ result \cite{EckerPR87} is recovered (dotted line in Fig.6b):
\begin{equation}
\mathcal{B}\left(  K_{S}\rightarrow\pi^{0}\gamma\gamma\right)  _{z>0.2}%
^{SU\left(  3\right)  ,\mathcal{O(}p^{4})}=3.8\times10^{-8}\;. \label{Ks9}%
\end{equation}

- Terms of order $m_{0}^{-2}$ introduce a class of $\mathcal{O}(p^{6})$
effects, leading in $1/N_{c}$, which corresponds exactly to the $L_{7}%
,N_{13},L_{9}^{(6)}$ and $N_{1}^{(6)}$ counterterms once these are saturated
by $\eta_{0}$ exchanges (see appendix, Eqs.(\ref{Ap4},\ref{Ap5},\ref{Ap7})).
They are seen to conspire collectively at the rate level (dashed line in
Fig.6b), already without $G_{8}^{s}$, in order to reproduce the effect of the
$\eta$ pole as a function of $m_{0}$ discussed before.

- Keeping the full $U(3)$ result comes to keeping some leading $1/N_{c}$
contributions at all orders. As the plain line in Fig.6b shows, the leading
$1/N_{c}$ terms of order $\mathcal{O}(p^{n>6})$ amount to only a small
correction to the rate (at least as long as $m_{0}$ is not too small).
\end{quote}

\subsection{The $K_{S}\rightarrow\pi^{0}\gamma\gamma$ rate and its sensitivity
to the penguin fraction}

One of the advantages of dealing with $\eta_{0}$ effects in the $U(3)$
framework rather than through $SU(3)$ local counterterms is that we do not
have to face the problem of fixing the $m_{0}$ parameter. Indeed, as discussed
in Sec.3.2, the requirement of freezing the pseudoscalar poles at their right
places, thereby restoring the analytical properties of the amplitude, leads to
a well-defined prescription. Since both $\hat{Q}_{1}$ and $\hat{Q}_{6}$
contribute at lowest order, Eq.(\ref{PMP3}) can be used directly, and we get
to%
\begin{align}
B_{8}\left(  T^{2}\right)   &  =\frac{3(2M_{K}^{2}-M_{\pi}^{2}-T^{2})}%
{8(M_{K}^{2}-M_{\pi}^{2})(T^{2}-M_{\pi}^{2})}-\frac{2M_{K}^{2}+M_{\pi}%
^{2}-3T^{2}}{8(M_{K}^{2}-M_{\pi}^{2})}\left(  \frac{c_{\theta}c_{\eta}}%
{T^{2}-M_{\eta}^{2}}+\frac{s_{\theta}c_{\eta^{\prime}}}{T^{2}-M_{\eta^{\prime
}}^{2}}\right) \nonumber\\
&  +\frac{1}{2\sqrt{2}}\left(  \frac{s_{\theta}c_{\eta}}{T^{2}-M_{\eta}^{2}%
}-\frac{c_{\theta}c_{\eta^{\prime}}}{T^{2}-M_{\eta^{\prime}}^{2}}\right)
\;,\label{Ks10}\\
B_{8}^{s}\left(  T^{2}\right)   &  =\frac{3}{4\sqrt{2}}\left(  \frac{s_{\theta
}c_{\eta}}{T^{2}-M_{\eta}^{2}}-\frac{c_{\theta}c_{\eta^{\prime}}}%
{T^{2}-M_{\eta^{\prime}}^{2}}\right)  \;,\nonumber\\
B_{27}\left(  T^{2}\right)   &  =\frac{-3(2M_{K}^{2}-M_{\pi}^{2}-T^{2}%
)}{8(M_{K}^{2}-M_{\pi}^{2})(T^{2}-M_{\pi}^{2})}+\frac{2M_{K}^{2}+M_{\pi}%
^{2}-3T^{2}}{8(M_{K}^{2}-M_{\pi}^{2})}\left(  \frac{c_{\theta}c_{\eta}}%
{T^{2}-M_{\eta}^{2}}+\frac{s_{\theta}c_{\eta^{\prime}}}{T^{2}-M_{\eta^{\prime
}}^{2}}\right)  \;,\nonumber
\end{align}
with $c_{\eta}\equiv c_{\theta}-2\sqrt{2}s_{\theta}$ and $c_{\eta^{\prime}%
}\equiv s_{\theta}+2\sqrt{2}c_{\theta}$, the mixing angle combinations for
$\eta$ and $\eta^{\prime}\rightarrow\gamma\gamma$, respectively (as a
consistency check, note that Eq.(\ref{Ks7}) is preserved).%
\begin{figure}
[t]
\begin{center}
\includegraphics[
height=1.7919in,
width=6.2275in
]%
{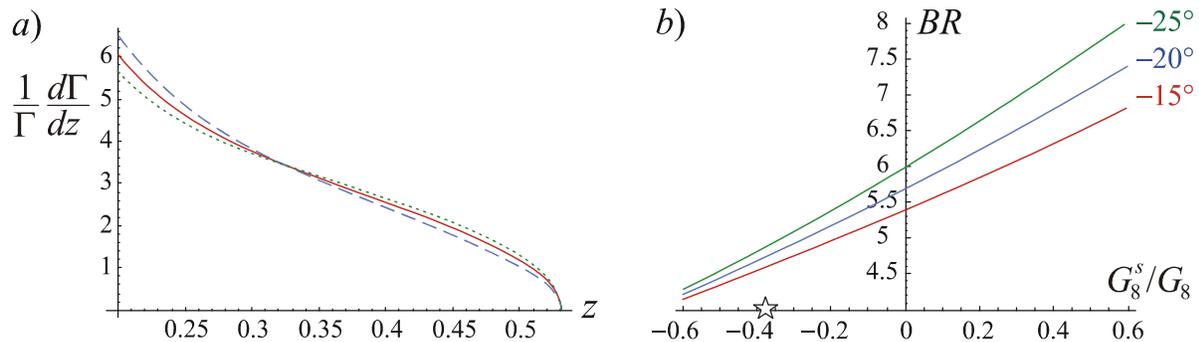}%
\caption{a) $K_{S}\rightarrow\pi^{0}\gamma\gamma$ normalized differential rate
for $\theta_{P}=-20{{}^\circ}$, $G_{8}^{s}/G_{8}=-0.3,0,+0.3$ (dashed, plain,
dotted). b) $\mathcal{B}\left(  K_{S}\rightarrow\pi^{0}\gamma\gamma\right)  $,
$\times$ $10^{8}$, as a function of $G_{8}^{s}/G_{8}$ for $\theta
_{P}=-15{{}^\circ},-20{{}^\circ},-25{{}^\circ}$. The star refers to
Eq.(\ref{QCDinspG8s}).}%
\end{center}
\end{figure}

The resulting differential rate for $z>0.2$ is shown in Fig.7a. Yet its shape
is not much affected by neither $\theta_{P}$ nor $G_{8}^{s}/G_{8}$, and
extracting information on the latter is clearly beyond foreseeable
experimental sensitivity. Looking at the low-energy end of the $\gamma\gamma$
spectrum would not be more helpful. Indeed, integrating over $0<z<0.048$
($m_{\gamma\gamma}<108$ MeV), the obtained partial branching ratio varies from
$0.48\times10^{-8}$ to $0.44\times10^{-8}$ for $G_{8}^{s}/G_{8}$ between
$-0.6$ and $0.6$, while the $SU(3)$ prediction is $0.49\times10^{-8}$.

Fortunately, for the total rate, the situation is better though, as said
before, $G_{8}^{s}$ is not dominant. Fig.7b summarizes the constraints one
could get on $G_{8}^{s}/G_{8}$ from an experimental determination of the
branching ratio. Note that merely fixing the sign of $G_{8}^{s}/G_{8}$
(inaccessible from $K_{L}\rightarrow\gamma\gamma$ alone) would already give
valuable information on the low-energy realization of the effective weak
Hamiltonian (\ref{EffWeakHam}).

Of course, Fig.7b is subject to the theoretical uncertainties associated with
any leading order ChPT computation. A precise extraction of $G_{8}^{s}/G_{8}$
from $K_{S}\rightarrow\pi^{0}\gamma\gamma$ would require an estimation of the
NLO effects. While some unitary corrections are already included through the
use of $G_{8}$ as extracted from $K_{S}\rightarrow\pi\pi$, there remains the
separate class of $\mathcal{O}(p^{6})$ effects from vector mesons. These were
seen to give sizeable contributions to $K_{L}\rightarrow\pi^{0}\gamma\gamma$
(see Ref.\cite{BDI03} for a recent review and list of references) and though
the underlying dynamics is very different for $K_{S}\rightarrow\pi^{0}%
\gamma\gamma$, working out their impact would be really called for. In
particular, one should investigate if they affect either the differential rate
(Fig.7a) or the total rate for low $\gamma\gamma$ invariant mass, which are
pretty much insensitive to $G_{8}^{s}/G_{8}$ and $\theta_{P}$. In that case,
these two observables would provide essential discriminating tools, able to
single out the $\mathcal{O}(p^{6})$ corrections due to vector resonances.

Finally, one can predict $\mathcal{B}\left(  K_{S}\rightarrow\pi^{0}%
\gamma\gamma\right)  $ from the range of $G_{8}^{s}/G_{8}$ preferred by
Eq.(\ref{LeadLog}) (which englobes the one extracted from $K_{L}%
\rightarrow\gamma\gamma$, see Eq.(\ref{G8sPhTh})):%
\begin{equation}
G_{8}^{s}/G_{8}=-0.38\pm0.12\Rightarrow\mathcal{B}\left(  K_{S}\rightarrow
\pi^{0}\gamma\gamma\right)  _{z>0.2}^{U\left(  3\right)  ,\mathcal{O}(p^{4}%
)}=\left(  4.8\pm0.5\right)  \times10^{-8}\;, \label{Ks11}%
\end{equation}
where the error, inferred from Fig.7b, includes the theoretical uncertainties
associated with the physical mass prescription (i.e., with a class of
$\mathcal{O}(p^{6})$ effects). As expected, the effect of the $\eta_{0}$ meson
increases the $\mathcal{O}(p^{4})$ $SU(3)$ prediction (\ref{Ks9}). Note that
both results are in agreement with the (still imprecise) experimental value
(\ref{Ks1}).

\section{The $K^{+}\rightarrow\pi^{+}\gamma\gamma$ decay}

Let us now turn to $K^{+}\rightarrow\pi^{+}\gamma\gamma$, which has already
received a lot of attention. Experimentally, the situation has been improved
recently\cite{E949} and is expected to get even better in the near future. On
the theoretical side, the analysis is slightly more involved than for
$K_{S}\rightarrow\pi^{0}\gamma\gamma$ as there are now both loop and pole
contributions at $\mathcal{O}(p^{4})$ \cite{EckerPR88}:%
\begin{equation}
\mathcal{A}^{\mu\nu}\left(  K^{+}\rightarrow\pi^{+}\gamma\gamma\right)
=\frac{\alpha}{4\pi}\left[  A\left(  T^{2}\right)  \left(  T^{2}g^{\mu\nu
}-2k_{1}^{\nu}k_{2}^{\mu}\right)  +B\left(  T^{2}\right)  \varepsilon^{\mu
\nu\rho\sigma}k_{1\rho}k_{2\sigma}\frac{{}}{{}}\right]  \;,
\end{equation}
with $A(T^{2})$, the $\pi^{\pm},K^{\pm}$ loop function, and $B(T^{2})$, the
$\pi^{0},\eta,\eta^{\prime}$ pole one. These two pieces do not interfere in
the rate since they produce the two photons in different $CP$ states:
\begin{equation}
\Gamma\left(  K^{+}\rightarrow\pi^{+}\gamma\gamma\right)  =\frac{\alpha
^{2}M_{K}^{5}}{16\left(  4\pi\right)  ^{5}}\int_{0.2}^{z_{\max}}%
dz\lambda^{1/2}\left(  1,z,r_{\pi}^{2}\right)  z^{2}\left[  4\left|  A\left(
z\right)  \right|  ^{2}+\left|  B\left(  z\right)  \right|  ^{2}\frac{{}}{{}%
}\right]  \;, \label{Kp2}%
\end{equation}
where the cut in $z$ is applied to remove the $K^{+}\rightarrow\pi^{+}\pi^{0}$
background. Though the loop contribution is larger than the pole one, this
latter piece should not be neglected. Indeed, as we will see, it is very
sensitive to the ratio $G_{8}^{s}/G_{8}$ thanks to the suppression of the
$\pi^{0}$ pole compared to the $\eta,\eta^{\prime}$ ones ($K^{+}\rightarrow
\pi^{+}\pi^{0}$ is a $\Delta I=3/2$ transition).

\subsection{Loop amplitude in SU(3) ChPT}

For the $A\left(  z\right)  $ amplitude, working with the enlarged symmetry
$U(3)$ would be superfluous as only charged particles circulate the loop. We
will thus take up the standard $\mathcal{O}(p^{4})$ $SU(3)$ result
\cite{EckerPR88}:%
\begin{align}
A\left(  z\right)   &  =G_{8}\left[  \left(  1+\frac{1}{z}-\frac{r_{\pi}^{2}%
}{z}+\delta_{27}^{\pi}\right)  \Phi\left(  z/r_{\pi}^{2}\right)  +\left(
1-\frac{1}{z}+\frac{r_{\pi}^{2}}{z}+\delta_{27}^{K}\right)  \Phi\left(
z\right)  -\hat{c}\right]  \;,\nonumber\\
\delta_{27}^{\pi}  &  =\frac{G_{27}}{G_{8}}\left(  \frac{13r_{\pi}^{2}}%
{3z}+\frac{7}{3z}-\frac{13}{3}\right)  ,\;\;\delta_{27}^{K}=\frac{G_{27}%
}{G_{8}}\left(  \frac{7r_{\pi}^{2}}{3z}+\frac{13}{3z}-\frac{13}{3}\right)  \;,
\label{Kp3}%
\end{align}
with the three-point loop functions%
\begin{equation}
\Phi\left(  a\right)  =\left\{
\begin{array}
[c]{ll}%
1-\dfrac{4}{a}\arcsin^{2}\sqrt{a/4} & 4/a>1\\
1+\dfrac{1}{a}\left(  \log\frac{1-\sqrt{1-4/a}}{1+\sqrt{1-4/a}}+i\pi\right)
^{2}\;\;\; & 0<4/a<1\;.
\end{array}
\right.  \label{Kp4}%
\end{equation}
The contributions of the counterterms are collected in the unknown constant
$\hat{c}$:
\begin{equation}
\hat{c}=\frac{128\pi^{2}}{3}\left(  3\left(  L_{9}+L_{10}\right)
+N_{14}-N_{15}-2N_{18}+\frac{2G_{27}}{3G_{8}}\left(  3\left(  L_{9}%
+L_{10}\right)  +D_{i}\right)  \right)  \,, \label{Kp5}%
\end{equation}
where the $L_{i}$'s refer to the basis of $\mathcal{O}(p^{4})$ strong
operators of Gasser-Leutwyler \cite{GasserL84} and the $N_{i}$'s to the octet
$|\Delta S|=1$ one of Ecker-Kambor-Wyler \cite{EckerKW93}. The 27-plet
counterterms, suppressed by the $\Delta I=1/2$ rule, are collectively denoted
by $D_{i}$ \cite{Esposito91,KamborMW90}. The $L_{i}$, $N_{i}$ and $D_{i}$
occur in separately finite combinations in $\hat{c}$. Finally, note that the
weak mass operator does not contribute. This has to be so since it could not
have been absorbed in the weak counterterms $N_{14,15,18}$.

The above loop amplitude induces the following branching fraction for $z>0.2$:%
\begin{equation}
\mathcal{B}\left(  K^{+}\rightarrow\pi^{+}\gamma\gamma\right)  _{z>0.2}%
^{\mathrm{L},SU(3),\mathcal{O}(p^{4})}=\left(
5.77+1.64\hat{c}+0.29\hat{c}^{2}\right)  \times10^{-7}\;. \label{Kp6}%
\end{equation}
This expression is not much affected by the momentum cut, as can be inferred
from the shape of the two-photon invariant mass spectrum which exhibits the
characteristic peak above the $\pi^{+}\pi^{-}$ threshold.

At $\mathcal{O}(p^{6})$, unitarity corrections from $K\rightarrow\pi\pi\pi$ as
well as vector meson effects must be included, and were analyzed in
Ref.\cite{DAmbrosioP96Kppgg}. The former increase the rate by about $30-40\%$
while the latter should be smaller. From the corrected spectrum and rate,
but\ assuming negligible pole contributions, a fit to the E787 data
\cite{E787}, (here we quote only the branching fraction)%
\begin{equation}
\mathcal{B}\left(  K^{+}\rightarrow\pi^{+}\gamma\gamma\right)
_{100\,\text{MeV}<P_{\pi^{+}}<180\,\text{MeV}}^{\exp}=\left(  6.0\pm
1.5\pm0.7\right)  \times10^{-7}\; \label{Kp7}%
\end{equation}
with $P_{\pi^{+}}=\frac{M_{K}}{2}\lambda\left(  1,r_{\pi}^{2},z\right)
^{1/2}$, the $\pi^{+}$ momentum, leads to the $\mathcal{O}(1)$ positive value%
\begin{equation}
\hat{c}=1.8\pm0.6. \label{c_hat}%
\end{equation}

\subsection{Pole amplitude in U(3) ChPT}

The pole diagrams for $K^{+}\rightarrow\pi^{+}\gamma\gamma$ are similar to
those for $K_{S}\rightarrow\pi^{0}\gamma\gamma$, with in addition a
five-particle contact interaction from the WZW term (\ref{WZW2}) in the case
of the weak mass operator $Q_{8}^{m}$ (Fig.8). A straightforward leading order
computation in $U(3)$ ChPT gives:%
\begin{figure}
[t]
\begin{center}
\includegraphics[
height=1.2471in,
width=5.8928in
]%
{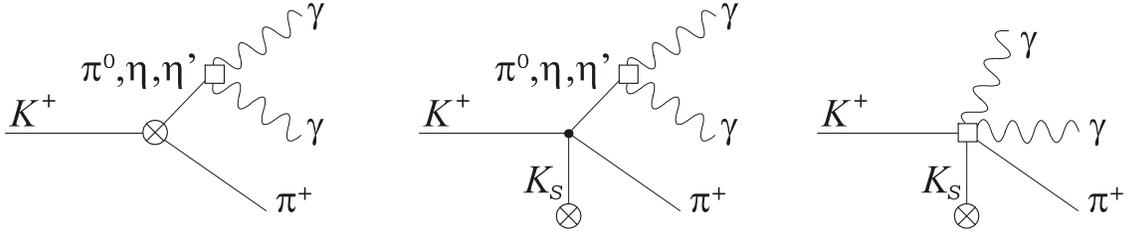}%
\caption{Pole and tadpole diagrams for the process $K^{+}\rightarrow\pi
^{+}\gamma\gamma$.}%
\end{center}
\end{figure}
\begin{equation}
B\left(  T^{2}\right)  =\frac{16}{3}(m_{K}^{2}-m_{\pi}^{2})\left[  G_{8}%
B_{8}\left(  T^{2}\right)  +G_{8}^{s}B_{8}^{s}\left(  T^{2}\right)
+G_{27}B_{27}\left(  T^{2}\right)  \right]  \,, \label{Kp8}%
\end{equation}
with%
\begin{gather}
B_{8}\left(  T^{2}\right)  =\frac{m_{0}^{2}-10m_{K}^{2}+3m_{\pi}^{2}+7T^{2}%
}{2(T^{2}-m_{\eta}^{2})(T^{2}-m_{\eta^{\prime}}^{2})}\,,\;B_{8}^{s}\left(
T^{2}\right)  =\frac{-5m_{K}^{2}+2m_{\pi}^{2}+3T^{2}}{(T^{2}-m_{\eta}%
^{2})(T^{2}-m_{\eta^{\prime}}^{2})}\,,\label{Kp9}\\
B_{27}\left(  T^{2}\right)  =\frac{m_{0}^{2}(5m_{K}^{2}-m_{\pi}^{2}%
-4T^{2})+(15m_{K}^{2}-7m_{\pi}^{2}-8T^{2})(m_{\pi}^{2}-T^{2})}{3\left(
T^{2}-m_{\pi}^{2}\right)  (T^{2}-m_{\eta}^{2})(T^{2}-m_{\eta^{\prime}}^{2}%
)}\,.\nonumber
\end{gather}
Again, $m_{\pi,K,\eta,\eta^{\prime}}$ stand for the theoretical masses, which
ensures the absence of $Q_{8}^{m}$ effects when the two tadpole graphs are
included. Unlike for $K_{S}\rightarrow\pi^{0}\gamma\gamma$, the three pole
functions $B_{8},B_{8}^{s}$ and $B_{27}$ are now independent as $\hat{Q}_{1}$,
$\hat{Q}_{2}$ and $\hat{Q}_{6}$ all contribute. Note that the pion pole
disappears in $B_{8}$, leaving a constant term, because of the $T^{2}-m_{\pi
}^{2}$ momentum dependence of the off-shell $K^{+}\rightarrow\pi^{+}\pi^{0}$
amplitude (purely $\Delta I=3/2$ once on-shell).

The pole-induced branching fraction as a function of $m_{0}$ is depicted in
Fig.9a. It is not a monotonically decreasing function down to the $SU(3)$
limit as in the $K_{S}\rightarrow\pi^{0}\gamma\gamma$ case. On the contrary,
it exhibits a strong suppression for $m_{0}$ between $1.0$ and $1.5 $ GeV.
Below 1 GeV, the $\eta$ pole begins to be felt significantly, hence the rate
increases as in Fig.6a.%
\begin{figure}
[t]
\begin{center}
\includegraphics[
height=1.727in,
width=6.3676in
]%
{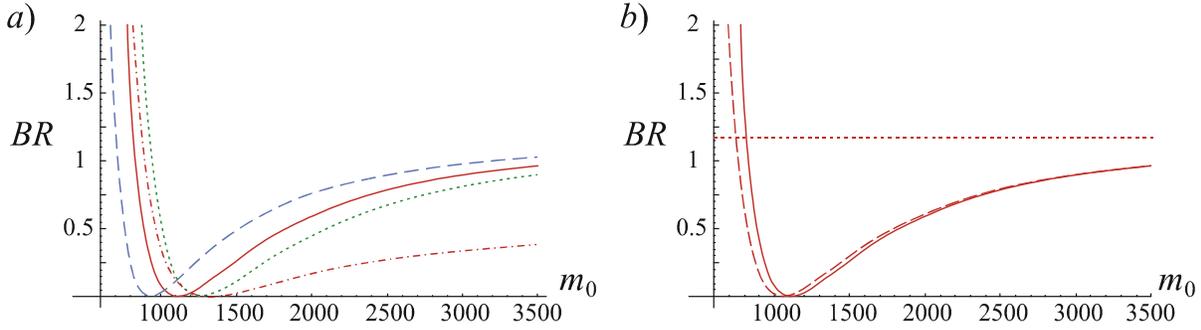}%
\caption{a) $\mathcal{B}\left(  K^{+}\rightarrow\pi^{+}\gamma\gamma\right)
^{poles}$, $\times$ $10^{7}$, as a function of $m_{0}$ for $G_{8}^{s}%
/G_{8}=-0.3,0,+0.3$ (dashed, plain, dotted) and for $G_{8}$ alone
(dash-dotted). b) Comparison between the $\mathcal{O}(p^{4})$ $SU(3)$ result
(dotted), idem plus the $m_{0}^{-2}$ corrections (dashed), and full
$\mathcal{O}(p^{4})$ $U(3)$ result (plain), with $G_{8}^{s}=0$ for the three
curves.}%
\end{center}
\end{figure}

\subsection{Reduction of the pole amplitude to SU(3) ChPT}

The $SU(3)$ limit of the above result can be investigated expanding the pole
functions in powers of $1/m_{0}^{2}$:%
\begin{equation}%
\begin{array}
[c]{cclc}%
B_{8}\left(  T^{2}\right)  = & \dfrac{-1/2}{T^{2}-m_{\eta8}^{2}} &
-\dfrac{4(5m_{K}^{2}-2m_{\pi}^{2}-3T^{2})(m_{K}^{2}-T^{2})}{3(T^{2}-m_{\eta
8}^{2})^{2}m_{0}^{2}} & +\mathcal{O}(m_{0}^{-4})\;,\\
B_{8}^{s}\left(  T^{2}\right)  = & 0 & +\dfrac{5m_{K}^{2}-2m_{\pi}^{2}-3T^{2}%
}{(T^{2}-m_{\eta8}^{2})m_{0}^{2}} & +\mathcal{O}(m_{0}^{-4})\;,\\
B_{27}\left(  T^{2}\right)  = & \dfrac{5/4}{T^{2}-m_{\pi}^{2}}+\dfrac
{1/12}{T^{2}-m_{\eta8}^{2}} & -\dfrac{2(5m_{K}^{2}-2m_{\pi}^{2}-3T^{2}%
)(3m_{K}^{2}-m_{\pi}^{2}-2T^{2})}{9(T^{2}-m_{\eta8}^{2})^{2}m_{0}^{2}} &
+\mathcal{O}(m_{0}^{-4})\;,
\end{array}
\label{Kp10}%
\end{equation}
The behavior of the total rate as a function of $m_{0}$ when the $m_{0}$
series is truncated at a given order is displayed in Fig.9b:

\begin{quote}
- The $\mathcal{O}(p^{4})$ $SU(3)$ amplitude (dotted line in Fig.9b)
corresponds to the $\mathcal{O}(m_{0}^{0})$ terms \cite{EckerPR88}, and gives%
\begin{equation}
\mathcal{B}\left(  K^{+}\rightarrow\pi^{+}\gamma\gamma\right)  _{z>0.2}%
^{\mathrm{P},SU(3),\mathcal{O}(p^{4})}=1.17\times10^{-7}\;. \label{Kp11}%
\end{equation}
This result is to be compared to $0.51\times10^{-7}$ without the $G_{27}$
piece. Such a large effect of the 27 operator (see Fig.9a) has been overlooked
previously, and arises from the pion pole, absent from the octet, that
compensates for the $\Delta I=1/2$ suppression over the whole phase-space.

- The $m_{0}^{-2}$ corrections are the collective effect of the $SU(3)$
counterterms $L_{7},N_{13},L_{9}^{(6)}$ and $N_{1}^{(6)}$, when saturated by
the $\eta_{0}$ (see appendix, Eqs.(\ref{Ap4},\ref{Ap5},\ref{Ap7})). The series
seems not well-behaved in the case of $B_{8}$, again because the leading term
is suppressed by the absence of the $\pi^{0}$ pole factor. In particular, for
$m_{0}$ between $1$ GeV and $1.5$ GeV, the $m_{0}^{-2}$ and $m_{0}^{0}$ terms
interfere destructively and even cancel each other out for a given $T^{2}$
inside the phase-space. This is the origin of the dip shown in Fig.9a, as
confirmed by the dashed line in Fig.9b. Said differently, the $\Delta I=1/2$
enhanced $B_{8}$ amplitude, being essentially due to the $\eta_{8}$ pole, is
very sensitive to $\eta-\eta^{\prime}$ mixing effects.

- It is a well-known fact that when the leading order is suppressed for some
reason, one can expect sizeable effects from the NLO corrections. Since in the
present case the NLO corrections are not suppressed by any mechanism (compare
the $m_{0}^{-2}$ correction in Eqs.(\ref{Kp10}) and (\ref{Ks7})), the series
is expected to behave correctly afterwards. The plain line in Fig.9b shows
that this is indeed the case: the full $U(3)$ result is rather well reproduced
by the $m_{0}^{-2}$ corrections alone.
\end{quote}

Note, finally, that for $m_{0}<1$ GeV, $B_{8}$ and $B_{27}$ begin to interfere
destructively, instead of constructively in the $SU(3)$ limit.

\subsection{Pole contribution to the rate}

Given the strong sensitivity of the pole amplitude to the $\eta-\eta^{\prime}$
system, the prescription Eq.(\ref{PMP3}) has to be enforced before numerical
evaluation. Expressions similar to those of Eq.(\ref{Ks10}) are then obtained:%
\begin{align}
B_{8}\left(  T^{2}\right)   &  =\frac{3}{4(M_{K}^{2}-M_{\pi}^{2}%
)}+\frac{2M_{K}^{2}+M_{\pi}^{2}-3T^{2}}{4(M_{K}^{2}-M_{\pi}^{2})}\left(
\frac{c_{\theta}c_{\eta}}{T^{2}-M_{\eta}^{2}}+\frac{s_{\theta}c_{\eta^{\prime
}}}{T^{2}-M_{\eta^{\prime}}^{2}}\right) \nonumber\\
&  -\frac{1}{\sqrt{2}}\left(  \frac{s_{\theta}c_{\eta}}{T^{2}-M_{\eta}^{2}%
}-\frac{c_{\theta}c_{\eta^{\prime}}}{T^{2}-M_{\eta^{\prime}}^{2}}\right)
\;,\label{Kp12}\\
B_{8}^{s}\left(  T^{2}\right)   &  =-\frac{3}{2\sqrt{2}}\left(
\frac{s_{\theta}c_{\eta}}{T^{2}-M_{\eta}^{2}}-\frac{c_{\theta}c_{\eta^{\prime
}}}{T^{2}-M_{\eta^{\prime}}^{2}}\right)  \;,\nonumber\\
B_{27}\left(  T^{2}\right)   &  =\frac{5M_{K}^{2}-7M_{\pi}^{2}+2T^{2}}%
{4(M_{K}^{2}-M_{\pi}^{2})(T^{2}-M_{\pi}^{2})}+\frac{3M_{K}^{2}-M_{\pi}%
^{2}-2T^{2}}{4(M_{K}^{2}-M_{\pi}^{2})}\left(  \frac{c_{\theta}c_{\eta}}%
{T^{2}-M_{\eta}^{2}}+\frac{s_{\theta}c_{\eta^{\prime}}}{T^{2}-M_{\eta^{\prime
}}^{2}}\right)  \;,\nonumber
\end{align}
with $c_{\eta}\equiv c_{\theta}-2\sqrt{2}s_{\theta}$ and $c_{\eta^{\prime}%
}\equiv s_{\theta}+2\sqrt{2}c_{\theta}$, as in Sec.4.3 \footnote{Note that,
for both $K_{S}\rightarrow\pi^{0}\gamma\gamma$ and $K^{+}\rightarrow\pi
^{+}\gamma\gamma$, the two-mixing angle result is immediately obtained by
subtituting ($f_{i}=\sec\left(  \theta_{0}-\theta_{8}\right)  F_{\pi}/F_{i}$)%
\begin{align*}
c_{\theta}c_{\eta^{\prime}}  &  \rightarrow f_{0}c_{8}(f_{8}s_{0}+2\sqrt
{2}f_{0}c_{8}),\;s_{\theta}c_{\eta^{\prime}}\rightarrow f_{8}s_{0}(f_{8}%
s_{0}+2\sqrt{2}f_{0}c_{8})\\
s_{\theta}c_{\eta}  &  \rightarrow f_{0}s_{8}(f_{8}c_{0}-2\sqrt{2}f_{0}%
s_{8}),\;c_{\theta}c_{\eta}\rightarrow f_{8}c_{0}(f_{8}c_{0}-2\sqrt{2}%
f_{0}s_{8})
\end{align*}
}. The pole contribution to the rate as a function of $G_{8}^{s}/G_{8}$ is
depicted in Fig.10b for $z>0.2$.

In the positive sign alternative for $G_{8}^{s}/G_{8}$, it could account for
an increase of the total rate by more than $20\%$, and should be taken into
account in the extraction of $\hat{c}$, Eq.(\ref{c_hat}). The shape of the
differential rate could be affected too, but only mildly, as illustrated for a
particular set of parameters in Fig.10a.%
\begin{figure}
[t]
\begin{center}
\includegraphics[
height=1.8697in,
width=6.3079in
]%
{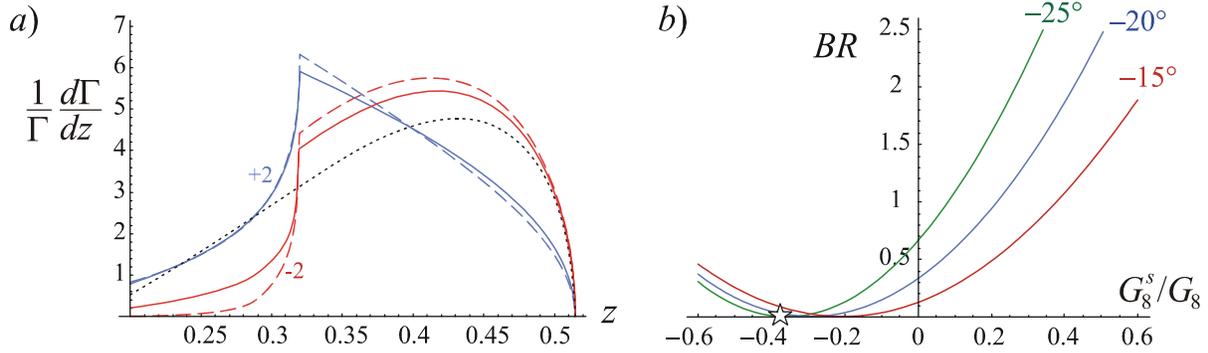}%
\caption{a) $K^{+}\rightarrow\pi^{+}\gamma\gamma$ normalized differential rate
for $\mathcal{O}(p^{4})$ loop + pole (plain), loop alone (dashed) and pole
alone (dotted) contributions, with $\hat{c}=\pm2$ and $G_{8}^{s}/G_{8}=+1/3,$
$\theta_{P}=-20{{}^\circ}$. b) $\mathcal{B}\left(  K^{+}\rightarrow\pi
^{+}\gamma\gamma\right)  ^{poles}$, $\times$ $10^{7}$, as a function of
$G_{8}^{s}/G_{8}$ for $\theta_{P}=-15{{}^\circ},-20{{}^\circ},-25{{}^\circ}$.
The star refers to Eq.(\ref{QCDinspG8s}).}%
\end{center}
\end{figure}

Conversely, for the preferred value $G_{8}^{s}/G_{8}\simeq-1/3$, the pole
contribution is completely suppressed (about $10^{-9}$). This is a pure
numerical coincidence: the three contributions $Q_{8,8s,27}$ are of the same
order, but can interfere destructively for $G_{8}^{s}/G_{8}<0$. No definite
prediction can be made in this case since $\mathcal{O}(p^{6})$ corrections can
no longer be neglected. Let us thus simply give an upper bound:%
\begin{equation}
G_{8}^{s}/G_{8}=-0.38\pm0.12\Rightarrow\mathcal{B}\left(  K^{+}\rightarrow
\pi^{+}\gamma\gamma\right)  _{z>0.2}^{\mathrm{P},U(3),\mathcal{O}(p^{4}%
)}<0.3\times10^{-7}\;. \label{Kp13}%
\end{equation}
This could go up to around $0.5\times10^{-7}$ with $\mathcal{O}(p^{6})$
effects but, in any case, no signal of pole contribution should show up
experimentally, neither in the total nor in the differential rate.

\subsection{Analysis of the low energy end of the $\gamma\gamma$ spectrum}

Recently, the E949 Collaboration has obtained the following upper limit
\cite{E949}:%
\begin{equation}
\mathcal{B}\left(  K^{+}\rightarrow\pi^{+}\gamma\gamma\right)  _{P_{\pi^{+}%
}>213\,\text{MeV}}^{\exp}<8.3\times10^{-9}\;,
\end{equation}
assuming a spectrum of the shape predicted by ChPT with unitarity corrections.
This corresponds to a $\gamma\gamma$ invariant mass in the range $0<z<0.0483$,
below the $\pi^{0}$ pole. The loop and pole leading order $SU(3)$ predictions
for this range are given by:%
\begin{align}
\mathcal{B}\left(  K^{+}\rightarrow\pi^{+}\gamma\gamma\right)  _{z<0.0483}%
^{\mathrm{L},SU(3),\mathcal{O}(p^{4})}  &  =\left(  0.06\hat{c}^{2}%
+0.15\hat{c}+0.09\right)  \times10^{-9}\;,\\
\mathcal{B}\left(  K^{+}\rightarrow\pi^{+}\gamma\gamma\right)  _{z<0.0483}%
^{\mathrm{P},SU(3),\mathcal{O}(p^{4})}  &  =1.0\times10^{-9}\;.
\label{BrKpPipggPSU3}%
\end{align}

For $\hat{c}$ of order $1$, the loop contribution should be $<10^{-9}$.
Including the unitarity corrections analyzed in Ref.\cite{DAmbrosioP96Kppgg},
which are specially large for small $z$, it raises to%
\begin{equation}
\mathcal{B}\left(  K^{+}\rightarrow\pi^{+}\gamma\gamma\right)  _{z<0.0483}%
^{\mathrm{L},SU(3),unitarized}=6.1\times10^{-9}\;.
\end{equation}
A large error (presumably more than 50\%) should be assigned to this number
given the rather poor theoretical control over this small corner of phase-space.%

\begin{figure}
[t]
\begin{center}
\includegraphics[
height=3.0675in,
width=3.6633in
]%
{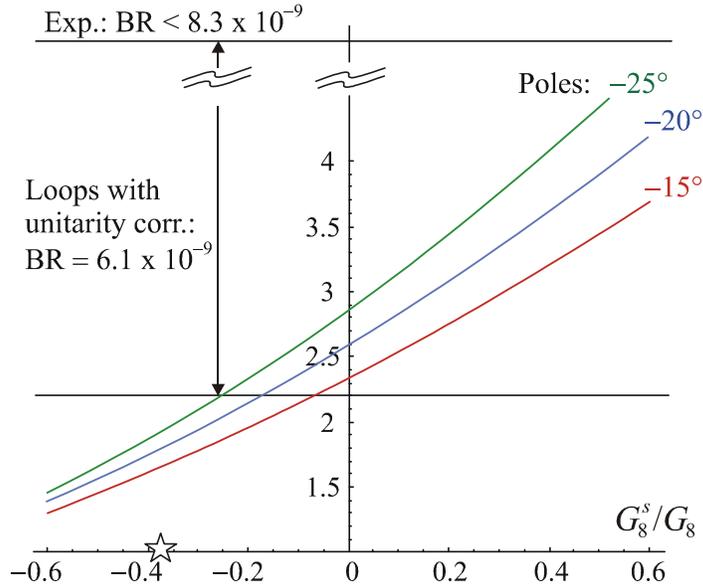}%
\caption{$\mathcal{B}\left(  K^{+}\rightarrow\pi^{+}\gamma\gamma\right)
^{poles}$ for $z<0.0483$, $\times\,10^{9}$. Assuming non-negligible loop
contributions \cite{DAmbrosioP96Kppgg}, the recent upper bound \cite{E949}
hints towards negative values for $G_{8}^{s}/G_{8}$. The star refers to
Eq.(\ref{QCDinspG8s}).}%
\end{center}
\end{figure}

The pole contribution, including $\eta_{0}$ effects, should also be taken into
account. Restricting the phase-space integration to the present range, we find
the situation shown in Fig.11. Interestingly, it appears that given the rather
large unitarity corrections, and the present experimental bound on the high
$\pi^{+}$ momentum end of the spectrum, negative values of $G_{8}^{s}/G_{8}$
are preferred. Together with the possibility of correlated study of the total
and differential rates, this makes of $K^{+}\rightarrow\pi^{+}\gamma\gamma$ a
promising mode to gain information on this ratio.

\section{Other radiative modes}

In this section, we consider two other modes involving pole amplitudes. Our
discussion will be very brief, because the sensitivity to the parameter of
interest $G_{8}^{s}/G_{8}$ will turn out to be either very suppressed like for
$K_{L}\rightarrow\pi^{0}\pi^{0}\gamma\gamma$, or buried among other unknown
parameters like for $K_{L}\rightarrow\pi^{+}\pi^{-}\gamma$.

\subsection{The $K_{L}\rightarrow\pi^{0}\pi^{0}\gamma\gamma$ decay}

This process is entirely due to pole diagrams at $\mathcal{O}(p^{4})$ (see
Fig.12), and was analyzed in $SU(3)$ ChPT in Ref.\cite{KLppgg}. The $U(3)$
amplitude is very cumbersome and will not be given explicitly. Its main
features are (independently of the use of the physical mass prescription):

\begin{quote}
- It receives contributions from $Q_{8},$ $Q_{8}^{s}$ and $Q_{27}$. As for
$K_{L}\rightarrow\gamma\gamma$, there is no tadpole graph but the contribution
of $Q_{8}^{m}$ still cancels out when the $\eta,\eta^{\prime} $ theoretical
masses are inserted.

- It depends on both the $T^{2}$ ($\gamma\gamma$ invariant mass) and $Q^{2} $
($\pi\pi$ invariant mass) kinematical variables, while the $SU(3)$ amplitude
depends only on $T^{2}$.

- As for $K_{S}\rightarrow\pi^{0}\gamma\gamma$, the $G_{8},G_{8}^{s}$ and
$G_{27}$ pieces are not independent because the current $\times$ current
$\hat{Q}_{2}$ operator does not contribute.%
\begin{figure}
[t]
\begin{center}
\includegraphics[
height=1.4356in,
width=4.2333in
]%
{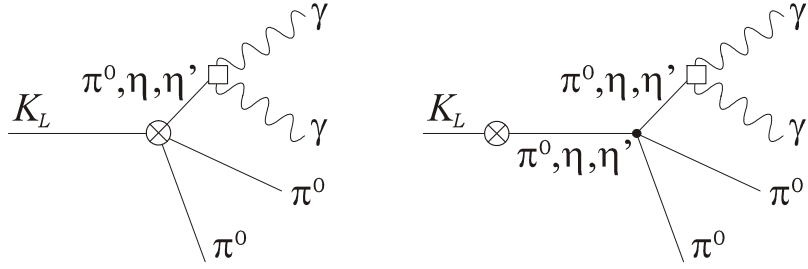}%
\caption{Typical pole diagrams for the process $K_{L}\rightarrow\pi^{0}\pi
^{0}\gamma\gamma$.}%
\end{center}
\end{figure}
\end{quote}

At first sight, one could think that the possible double occurrence of the
pseudoscalar poles would give a good sensitivity to $G_{8}^{s}$.
Unfortunately, this is not the case because the available phase-space for the
$\gamma\gamma$ invariant mass is very much concentrated around the $\pi^{0}$
pole. Since $\eta,\eta^{\prime}\rightarrow\pi^{0}\pi^{0}\pi^{0}$ are both
isospin violating, it is then again the $\pi^{0}$ pole that dominates the
initial $K_{L}\rightarrow P$ transition. The effects of the $\eta,\eta
^{\prime}$ and $G_{8}^{s}$ are thus completely suppressed.

Numerically, we have checked that, indeed, dealing with the mode in $U(3)$
ChPT with the physical mass prescription Eq.(\ref{PMP3}) changes the $SU(3)$
result of Ref.\cite{KLppgg} only at the percent level, no matter the
$\gamma\gamma$ invariant mass cuts. In addition, differential rates with
respect to either $T^{2}$ or $Q^{2}$ are only slightly modified. In
conclusion, the effects of the $\eta_{0}$ and $G_{8}^{s}$ are at the percent
level, beyond experimental sensitivity, and also beyond theoretical control
since $\mathcal{O}(p^{6})$ corrections to the dominant $\pi^{0}$ pole
amplitudes are certainly bigger than a few percents.

\subsection{The $K_{L}\rightarrow\pi^{+}\pi^{-}\gamma$ decay}

This mode receives many types of contributions, so let us briefly describe its
structure. The decay amplitude can be decomposed into electric and magnetic
transitions. The electric piece starts at $\mathcal{O}(p^{2})$, and is
completely determined at this order by the bremsstrahlung from $K_{L}%
\rightarrow\pi^{+}\pi^{-}$. It is therefore $CP$-violating and suppressed. At
$\mathcal{O}(p^{4})$, there is also a loop contribution, which is quite small
\cite{EckerNP94}.

This has allowed experimental access to the magnetic transition
\cite{KppgE731,KppgKTEV}. This one starts at $\mathcal{O}(p^{4})$ with
contributions from local $\Delta S=1$ odd-parity operators ($\sim
N_{29}+N_{31},$ see Ref.\cite{EckerNP94}) and pole diagrams (Fig.13a). In
addition, $\mathcal{O}(p^{6})$ effects were seen to be important from the
experimental observation of a non-zero slope in the photon energy. A detailed
study at that order is done in Ref.\cite{DAmbrosioP97Klppg}, including both
vector meson exchanges and chiral loops, and we will rely on that work for
conventions and numerics (as well as for a complete reference list).

In the present work, we wish only to comment on the pole part of the magnetic
amplitude (the $F_{1}$ piece in Refs.\cite{EckerNP94,DAmbrosioP97Klppg}, shown
in Fig.13a), which has been seen to be very sensitive to nonet symmetry
breaking \cite{Cheng90}. In our framework, this translates into a sensitivity
to $G_{8}^{s}$.

As for $K_{L}\rightarrow\gamma\gamma$, the pole amplitude vanishes in $SU(3)$
at $\mathcal{O}(p^{4})$ when enforcing the GMO relation. In $U(3)$, it is
simply obtained by substituting $V_{\gamma\gamma}^{\mu\nu}$ in Eq.(\ref{KL6})
by
\begin{equation}
V_{\pi^{+}\pi^{-}\gamma}^{\mu}=-\frac{e}{2\sqrt{2}\pi^{2}F^{3}}\left(
\begin{array}
[c]{c}%
1\\
0\\
0
\end{array}
\right)  i\varepsilon^{\mu\nu\rho\sigma}p_{\nu}^{\pi^{+}}p_{\rho}^{\pi^{-}%
}k_{\sigma}^{\gamma}\,, \label{Kg1}%
\end{equation}
and thus arises again only through the $\bar{u}u$ component of the weak
vertex, i.e. through $\hat{Q}_{1}$. As for $K_{L}\rightarrow\gamma\gamma$, NLO
effects from $\hat{Q}_{6}$ are assumed to behave according to usual chiral
counting and thus are discarded. Note also that the analysis of $K_{L}%
\rightarrow\gamma\gamma$ at $\mathcal{O}(p^{6})$ in $SU(3)$ presented in the
appendix would not be much modified for the present mode, and that its
conclusions are still valid. For these reasons, the phenomenological pole
parametrization of $F_{1}$ usually adopted is not appropriate and should be
replaced by%
\begin{equation}
F_{1}=-\frac{G_{8}^{s}+\frac{2}{3}G_{27}}{G_{8}}M_{K}^{2}\left(
\frac{1}{M_{K}^{2}-M_{\pi}^{2}}+\frac{(c_{\theta}-\sqrt{2}s_{\theta})^{2}%
}{3(M_{K}^{2}-M_{\eta}^{2})}+\frac{(s_{\theta}+\sqrt{2}c_{\theta})^{2}%
}{3(M_{K}^{2}-M_{\eta^{\prime}}^{2})}\right)  \;, \label{Kg2}%
\end{equation}
which has no $G_{8}$ contribution, is now dominated by the $\eta$ pole and is
thus negative when $G_{8}^{s}/G_{8}<0$.%
\begin{figure}
[t]
\begin{center}
\includegraphics[
height=1.8715in,
width=5.866in
]%
{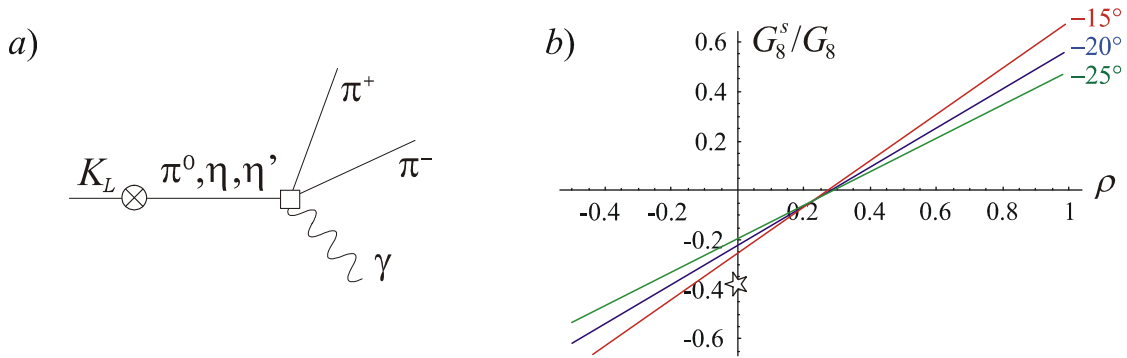}%
\caption{a) Anomalous pole contribution to the magnetic $K_{L}\rightarrow
\pi^{+}\pi^{-}\gamma$ amplitude. b) Numerical correspondance between the
parameter $\rho$ defined in \cite{DAmbrosioP97Klppg} and $G_{8}^{s}/G_{8}$
(see text). The star refers to Eq.(\ref{QCDinspG8s}).}%
\end{center}
\end{figure}

Instead of repeating the numerical analysis of Ref.\cite{DAmbrosioP97Klppg},
we have chosen to simply give a dictionary relating their parameter $\rho$ to
the ratio $G_{8}^{s}/G_{8}$ \textit{for fixed} $F_{1}$ (Fig.13b). In other
words, to each value of $F_{1}$ corresponds a given $\rho$ in
Ref.\cite{DAmbrosioP97Klppg}, and a given $G_{8}^{s}$ in Eq.(\ref{Kg2}), which
are then reported in Fig.13b. It is important to understand that this is only
a numerical equivalence, and does not correspond to the analytic equivalence
of Eq.(\ref{KL15}) (for example, $G_{8}^{s}/G_{8}=-1/3$ implies $\rho
\simeq0.5$ via Eq.(\ref{KL15}), but is accounted for by the smaller effective
value $\rho=-0.15\pm0.07$ in Fig.13b). As already explained in Sec.3.4, the
usual phenomenological pole model is incorrect because not proportional to
$\rho-1$ (to a good approximation).

Unfortunately, in Ref.\cite{DAmbrosioP97Klppg}, $\rho$ was only marginally
allowed into negative territory because this was believed to be in
contradiction with $K_{L}\rightarrow\gamma\gamma$. As we have seen,
$K_{L}\rightarrow\gamma\gamma$ points towards $G_{8}^{s}/G_{8}\simeq-1/3$, and
therefore negative values of $\rho$ should be allowed. From the fits of
Ref.\cite{DAmbrosioP97Klppg}, it seems that such values could accommodate the
data, but additional work would be needed. In any case, it is clear that a
precise extraction of $G_{8}^{s}/G_{8}$ from $K_{L}\rightarrow\pi^{+}\pi
^{-}\gamma$ would be quite intricate due to the presence of many counterterms
and vector meson couplings whose estimations introduce some amount of model dependence.

\section{Compatibility with hadronic observables}

This final section concerns non-radiative observables. First, we will see how
our understanding of $K_{L}\rightarrow\gamma\gamma$ can help estimating the
$\mathcal{O}(G_{F}^{2})$ $K_{L}-K_{S}$ mass difference generated by
pseudoscalar pole contributions. Then, we will turn to the direct CP-violating
parameter $\varepsilon^{\prime}/\varepsilon$, and analyze how the information
gained on the $\Delta I=1/2$ rule fits in the usual theoretical analysis.

\subsection{Pole contributions to $\Delta M_{LS}$}

Experimentally, the $K_{L}-K_{S}$ mass difference, $\Delta M_{LS}\equiv
M\left(  K_{L}\right)  -M\left(  K_{S}\right)  $, is quite
well-known\cite{PDG04}%
\begin{equation}
\Delta M_{LS}^{\exp}=\left(  3.483\pm0.006\right)  \times10^{-12}%
\;\text{MeV}\;. \label{DM1}%
\end{equation}
Unfortunately, theoretical control on $\Delta M_{LS}\sim2\operatorname{Re}%
\left\langle K^{0}\left|  \mathcal{H}_{W}\right|  \bar{K}^{0}\right\rangle $
is a long-standing issue since both short-distance and long-distance
contributions are present\cite{Wolfenstein79}%
\begin{equation}
\Delta M_{LS}=\Delta M_{LS}^{SD}+\Delta M_{LS}^{LD}\;. \label{DM2}%
\end{equation}
The first piece, from $W$-box diagrams with virtual $c,t$ quarks, accounts for
the bulk of the experimental value, $\Delta M_{LS}^{SD}/\Delta M_{LS}^{\exp
}=\left(  86\pm26\right)  \%$ \cite{HerrlichN95} (see also \cite{CataP04}).
The second one arises from $W$-box diagrams with low-virtuality $u$ quarks,
hence has to be dealt with light meson exchanges. In the present work, we will
concentrate on the long-distance, non-local contributions from double
insertion of $|\Delta S|=1$ transitions (see for example
\cite{Wolfenstein79,DeltaMLD}).

In $SU(3)$ ChPT, only pole and tadpole diagrams (Fig.14a) occur at
$\mathcal{O}(p^{2})$, giving rise to the $\mathcal{O}(G_{F}^{2})$ mass
difference:%
\begin{equation}
\Delta M_{LS}^{pole}\overset{SU(3)}{=}2F^{4}m_{K}^{3}\left(  G_{27}%
-G_{8}+G_{8}^{m}\right)  ^{2}\left(  \frac{1}{m_{K}^{2}-m_{\pi}^{2}%
}+\frac{1/3}{m_{K}^{2}-m_{\eta8}^{2}}\right)  \;. \label{DM3}%
\end{equation}
Though both $M\left(  K_{L}\right)  $ and $M\left(  K_{S}\right)  $ are
renormalized by a term in ($G_{8}^{m}$)$^{2}$ from the tadpole diagrams
(Fig.14a)\footnote{In the usual language (see Ref.\cite{WeakMassTerm}), this
is a manifestation of the necessary realignment of the vacuum brought by the
tadpole operator $Q_{8}^{m}$. As a result, at $\mathcal{O}(G_{F}^{2})$, the
physical $K_{L,S}$ masses are free parameters, and only the mass difference is
calculable.}, the mass difference is independent of $G_{8}^{m}$ as can be seen
by enforcing the GMO relation (\ref{KL3}). Still, everything cancels along
with it, and $\Delta M_{LS}^{LD}$ exactly vanishes.

Pseudoscalar poles thus begin to contribute at $\mathcal{O}(p^{4})$ in $SU(3)
$, through loop corrections to the masses and weak vertices, along with
genuine two-pseudoscalar loops, $K\rightarrow PP\rightarrow\bar{K}$, from
which they cannot be disentangled. Overall, these $\mathcal{O}(p^{4})$ loops
require some unknown $\mathcal{O}(p^{4})$ local counterterms, including
$|\Delta S|=2$ ones, to cancel their divergence (see the comment at the end of
the appendix). Alternatively, as they correspond to the low-energy tail of the
$W$-box diagrams, an approximate matching with short-distance can be
implemented \cite{BijnensGK91}.%
\begin{figure}
[t]
\begin{center}
\includegraphics[
height=3.1194in,
width=6.2111in
]%
{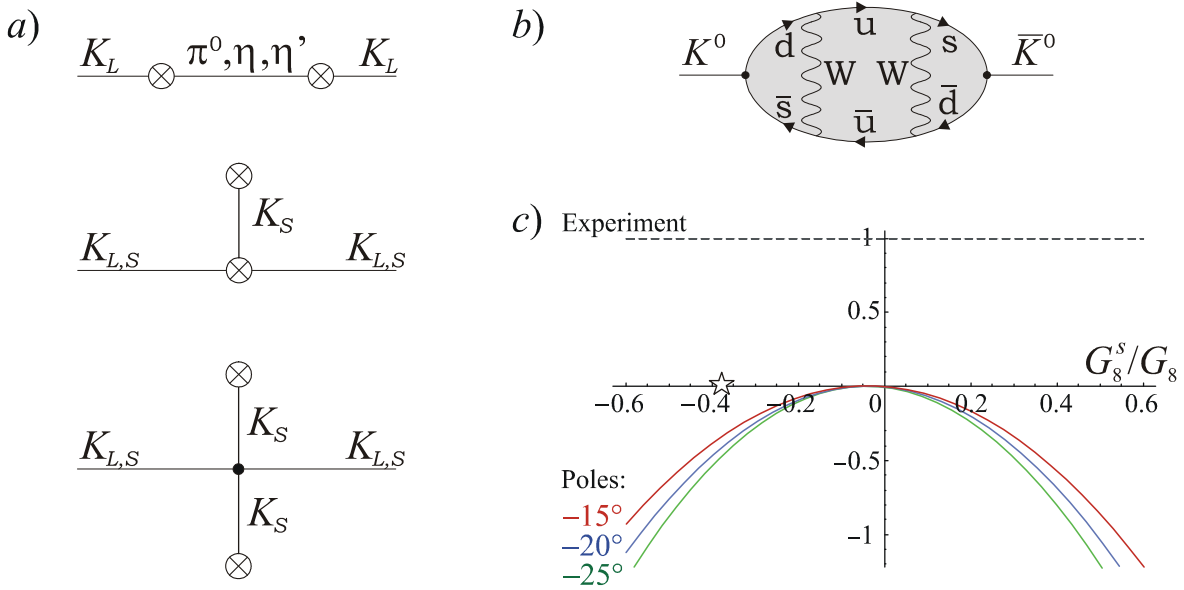}%
\caption{a) Pole and tadpole diagrams for $\Delta M_{LS}^{LD}$. b)
Long-distance $\bar{u}u$ contribution. c) Fraction of pole contribution to
$\Delta M_{LS}^{\exp}$ as a function of $G_{8}^{s}/G_{8}$ for $\theta
_{P}=-15{{}^\circ}$, $-20{{}^\circ}$, $-25{{}^\circ}$. The star refers to
Eq.(\ref{QCDinspG8s}).}%
\end{center}
\end{figure}

Performing the same analysis in $U(3)$ has the immediate advantage that the
transition $s\bar{d}\rightarrow\bar{u}u\rightarrow\bar{s}d$ is caught at
leading order (Fig.14b). Adapting Eq.(\ref{KL6}), with $V_{\gamma\gamma}%
^{\mu\nu}$ replaced by $V_{weak}$, we reach%
\begin{equation}
\Delta M_{LS}^{pole}\overset{U(3)}{=}\frac{-12F^{4}m_{K}^{3}}{m_{0}^{2}%
-3m_{K}^{2}+3m_{\pi}^{2}}\left(  G_{8}^{s}+\frac{2}{3}G_{27}\right)
^{2}=\frac{-12F^{4}m_{K}^{3}}{m_{0}^{2}-3m_{K}^{2}+3m_{\pi}^{2}}\left(
G_{W}x_{1}\right)  ^{2}\;. \label{DM4}%
\end{equation}
Though $Q_{8}$ and $Q_{8}^{m}$ cancel again upon enforcing the theoretical
$\eta_{0}-\eta_{8}$ mass matrix (\ref{KL5}), the non-zero contribution from
$(\hat{Q}_{1})^{2}$ survives (neglecting $\hat{Q}_{3}$ and $\hat{Q}_{5}$). As
for $K_{L}\rightarrow\gamma\gamma$, $\hat{Q}_{2}$ and $\hat{Q}_{6}$ cannot
contribute at leading order (such a disappearance of $\hat{Q}_{6}$ has already
been noticed in Ref.\cite{BurasGerard86}). The physical mass prescription is
then used to restore correct analytical properties, but only for the $(\hat
{Q}_{1})^{2}$ contribution%
\begin{equation}
\Delta M_{LS}^{pole}=2F^{4}M_{K}^{3}\left(  G_{8}^{s}+\frac{2}{3}%
G_{27}\right)  ^{2}\left(  \frac{1}{M_{K}^{2}-M_{\pi}^{2}}+\frac{(c_{\theta
}-\sqrt{2}s_{\theta})^{2}}{3(M_{K}^{2}-M_{\eta}^{2})}+\frac{(s_{\theta}%
+\sqrt{2}c_{\theta})^{2}}{3(M_{K}^{2}-M_{\eta^{\prime}}^{2})}\right)  \;.
\label{DM5}%
\end{equation}
Once again, the $\eta$ contribution dominates such that poles give a negative
contribution to $\Delta M_{LS}$, which grows (quasi) quadratically with
$G_{8}^{s}/G_{8}$ (Fig.14c). Note also the similarity of Eq.(\ref{DM5}) with
Eq.(\ref{Kg2}), stemming from the proportionality of the vertices $V_{\pi
^{+}\pi^{-}\gamma}^{\mu}$ (\ref{Kg1}) and $V_{weak}$ (\ref{KL9}) when
$x_{6}\rightarrow0$.

To reach Eq.(\ref{DM5}), we have discarded $x_{6}$ because it does not occur
at lowest order, i.e. in Eq.(\ref{DM4}). As explained in Sec. 3, in such
cases, the physical mass prescription would generate some contributions
corresponding to the inclusion of only a small class of higher order effects,
and pole amplitudes being plagued by cancellations, this is a very dangerous
procedure. In the case of the $K_{L}\rightarrow\gamma\gamma$ (and
$K_{L}\rightarrow\pi^{+}\pi^{-}\gamma$) radiative mode, we went on by arguing
that $\hat{Q}_{6}$ is in fact suppressed at higher order because the quantum
numbers of the initial $K_{L}$ together with the electromagnetic WZW vertex
project out non-$\bar{u}u$ transitions, and kept only the $\hat{Q}_{1}$
contribution in our comparison with experiment.

This last step cannot be extended to $\Delta M_{LS}$ since there is no such
projection from CP-symmetry and charge conservation for the $K^{0}-\bar{K}%
^{0}$ transition. In particular, the $(\hat{Q}_{6})^{2}$ contribution at
$\mathcal{O}(p^{4})$ is certainly not small now since it contains part of the
$K\rightarrow PP\rightarrow\bar{K}$ loops, i.e. the low-energy tail of the box
diagram depicted in Ref.\cite{BijnensGK91}. These loops give a positive
contribution to $\Delta M_{LS}$, and though their estimates vary greatly, they
are generally of a few tens of percent of $\Delta M_{LS}^{\exp}$.

Eq.(\ref{DM5}) is therefore the correct leading order estimation for the
long-distance $(\hat{Q}_{1})^{2}$ piece only. To compare it with experiment,
let us take, with respect to $\Delta M_{LS}^{\exp}$, the conservative upper
bound of $+50\%$ for the $K\rightarrow PP\rightarrow\bar{K}$ loops, assuming
they saturate the $(\hat{Q}_{6})^{2}$ contribution, and $+150\%$ for the
short-distance piece. We can then safely infer from Fig.14c that pole
contributions should be no less than $-100\%$, i.e.
\begin{equation}
|G_{8}^{s}/G_{8}|<0.6\;. \label{DM6}%
\end{equation}
Furthermore, for our preferred value $G_{8}^{s}/G_{8}\simeq-1/3$, poles
contribute for about $-30\%$ such that the long-distance contributions
partially cancel each other.

\subsection{Strong penguin contribution to $\varepsilon^{\prime}/\varepsilon$}

The direct CP-violation observable $\varepsilon^{\prime}/\varepsilon$ is
related to the imaginary part of the $K\rightarrow\pi\pi$ isospin amplitudes,
and additional theoretical inputs are necessary. We do not intend to give a
full account of $\varepsilon^{\prime}/\varepsilon$ theory (for recent
theoretical updates and references, see \cite{EpsiReview}), but would like to
see if the large penguin contribution to the $\Delta I=1/2$ rule found at the
hadronic scale is compatible with the small $\varepsilon^{\prime}/\varepsilon$
observed \cite{KTeV,NA48,NA31,E731}%
\begin{equation}
\left(  \frac{\varepsilon^{\prime}}{\varepsilon}\right)  _{\exp}=\left(
1.67\pm0.16\right)  \times10^{-3}\,. \label{Epsi1}%
\end{equation}

The general formula for this quantity is, in terms of the isospin amplitudes
defined in Eq.(\ref{Eq16}),%
\begin{equation}
\frac{\varepsilon^{\prime}}{\varepsilon}=e^{i\Phi}\frac{\omega}{\sqrt
{2}\left|  \varepsilon\right|  }\left[  \frac{\operatorname{Im}A_{2}%
}{\operatorname{Re}A_{2}}-\frac{\operatorname{Im}A_{0}}{\operatorname{Re}%
A_{0}}\right]  \;\;\;\text{with}\;\Phi=-\delta+\dfrac{\pi}{4}\approx0\;.
\label{Epsi2}%
\end{equation}
We will concentrate on the strong QCD penguin contribution, and thus discard
isospin breaking effects due to $\pi^{0}-\eta^{(\prime)}$ and electroweak
penguins. Keeping then only the $\operatorname{Im}A_{0}/\operatorname{Re}%
A_{0}$ part, we get in terms of matrix elements of the dominant $Q_{1},Q_{2}$
and $Q_{6}$ four-quark operators%
\begin{equation}
\left(  \frac{\varepsilon^{\prime}}{\varepsilon}\right)  _{0}=\frac{\omega
}{\sqrt{2}\left|  \varepsilon\right|  }\;\frac{\operatorname{Im}\lambda_{t}%
}{\operatorname{Re}\lambda_{u}}\;\frac{3y_{6}\left(  \mu\right)  \left\langle
Q_{6}\left(  \mu\right)  \right\rangle _{0}}{-z_{1}\left(  \mu\right)
\left\langle Q_{1}\left(  \mu\right)  \right\rangle _{0}+2z_{2}\left(
\mu\right)  \left\langle Q_{2}\left(  \mu\right)  \right\rangle _{0}%
+3z_{6}\left(  \mu\right)  \left\langle Q_{6}\left(  \mu\right)  \right\rangle
_{0}}\;, \label{Epsi3}%
\end{equation}
with $\left\langle Q_{i}\right\rangle _{0}\equiv\left\langle \pi\pi\left(
I=0\right)  \left|  Q_{i}\right|  K\right\rangle $ and $\lambda_{i}\equiv
V_{id}V_{is}^{\ast}$.

At this level, the usual treatment consists in taking the experimental value
for the denominator (i.e. $\operatorname{Re}A_{0}$), and trying to modelize
the $Q_{6}\left(  \mu\right)  $ matrix element in the numerator at a scale for
which $y_{6}\left(  \mu\right)  $ is calculable. In this respect, the penguin
fraction $\mathcal{F}_{P}$ does not help much since running it up to $\mu>1$
GeV throughout the non-perturbative regime is beyond our reach. Alternatively,
to make use of the extra theoretical and phenomenological information
$\mathcal{F}_{P}\simeq2/3$, see Fig.4, let us rewrite Eq.(\ref{Epsi3}) in a
way independent of the hadronic matrix elements \cite{GilmanWise}%
\begin{equation}
\left(  \frac{\varepsilon^{\prime}}{\varepsilon}\right)  _{0}=\frac{\omega
}{\sqrt{2}\left|  \varepsilon\right|  }\,\frac{\operatorname{Im}\lambda_{t}%
}{\operatorname{Re}\lambda_{u}}\;\mathcal{F}_{P}\;\frac{y_{6}\left(
\mu_{hadr}\right)  }{z_{6}\left(  \mu_{hadr}\right)  } \label{Epsi4}%
\end{equation}
with $\mu_{hadr}$ the typical scale of ChPT. The difficulty now is shifted
from getting up to $\left\langle Q_{6}\left(  \mu\right)  \right\rangle $ in
the perturbative regime, to getting down to the ratio $y_{6}\left(
\mu\right)  /z_{6}\left(  \mu\right)  $ in the non-perturbative regime.%
\begin{figure}
[t]
\begin{center}
\includegraphics[
height=1.727in,
width=3.0666in
]%
{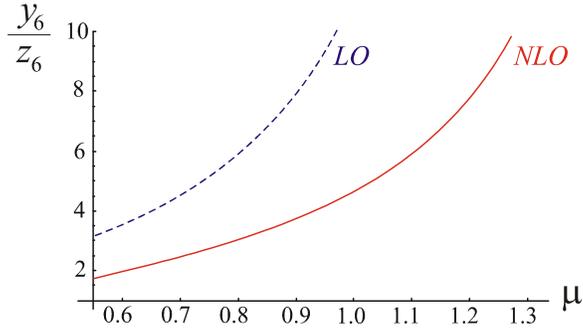}%
\caption{The running of the ratio $y_{6}(\mu)/z_{6}(\mu)$ at leading and
next-to-leading logarithmic approximation, in the $\overline{MS}$ / NDR
scheme, for $\alpha_{S}(M_{Z})=0.118$, $m_{c}=1.3$ GeV (based on
\cite{BuchallaBL96}).}%
\end{center}
\end{figure}

From the experimental constraints on the CKM factors $\lambda_{i}$, it follows%
\begin{equation}
\left(  \frac{\varepsilon^{\prime}}{\varepsilon}\right)  _{0}\simeq
\;5\,\mathcal{F}_{P}\;\frac{y_{6}\left(  \mu_{hadr}\right)  }{z_{6}\left(
\mu_{hadr}\right)  }\;\left(  \frac{\varepsilon^{\prime}}{\varepsilon}\right)
_{\exp}\;. \label{Epsi4b}%
\end{equation}
Allowing for at most a 50\% reduction due to isospin breaking effects, we
obtain then%
\begin{equation}
\frac{1}{5}\lesssim\mathcal{F}_{P}\;\frac{y_{6}\left(  \mu_{hadr}\right)
}{z_{6}\left(  \mu_{hadr}\right)  }\lesssim\frac{2}{5}\;. \label{Epsi5}%
\end{equation}
For $\mathcal{F}_{P}\simeq2/3$, this range is by no means unrealistic. As
shown in Fig.15, the ratio $y_{6}\left(  \mu\right)  /z_{6}\left(  \mu\right)
$ exhibits a scale dependence (this is obvious since $z_{6}\left(
m_{c}\right)  =0$ at LO), which quickly decreases with $\mu$
\cite{BuchallaBL96}. This behavior arises because at both LO and NLO in the
NDR scheme, $y_{6}\left(  \mu\right)  -z_{6}\left(  \mu\right)  $ is scale
independent to better than $10\%$ for $\mu$ even as low as $0.6$ GeV where
clearly perturbation theory should no longer be trusted. It is therefore
possible that though $y_{6}\left(  \mu_{hadr}\right)  $ and $z_{6}\left(
\mu_{hadr}\right)  $ are completely beyond our reach, their ratio could be
controlled. If, for once in the long story of $\varepsilon^{\prime
}/\varepsilon$, nature was kind enough to somehow protect the specific ratio
$y_{6}\left(  \mu\right)  /z_{6}\left(  \mu\right)  $, an alternative
theoretical strategy would be revived. This remains to be seen though all we
can say (on the basis of Fig.15) is that a value smaller than one for this
ratio is certainly not ruled out, and would require the bound%
\begin{equation}
G_{8}^{s}/G_{8}>-0.8\,, \label{Espi6}%
\end{equation}
which is not much more constraining than what would have been obtained from
the QCD-inspired analysis of Sec.2.3 (see Fig.1), i.e. $G_{8}^{s}/G_{8}>-1$.

\section{Conclusion}

A number of radiative $K$ decays involving pseudoscalar pole diagrams have
been investigated in the context of large-$N_{c}$ ChPT. Emphasis has been laid
on the $\Delta S=1$ weak operator $Q_{8}^{s}$, peculiar to the $U(3)$
framework, that holds the key to a better understanding of the underlying
flavor-changing mechanisms and, in particular, to a phenomenological
extraction of the penguin contribution to the $\Delta I=1/2$ rule. Let us now
summarize our results, divided into three categories:\bigskip

\textit{Phenomenological constraints on the weak coupling }$G_{8}^{s}$
\textit{from radiative }$K$\textit{\ decays}

\begin{itemize}
\item[--] The $K_{L}\rightarrow\gamma\gamma$ decay turns out to be mainly
driven by the nonet-symmetry breaking operators $Q_{8}^{s}$ and $Q_{27}$
(i.e., $\hat{Q}_{1}$), and dominated by the $\eta$ pole. The contribution of
$Q_{8}$ (i.e., $\hat{Q}_{6}$) is suppressed by large cancellations.
Experimental data then imply $G_{8}^{s}/G_{8}\simeq\pm1/3$. The analysis of
$K_{L}\rightarrow\gamma\gamma^{\ast}$ of Ref.\cite{DAmbrosioP97KLgg} favors
the negative solution. This sign is important for the interference between the
short-distance and dispersive $\gamma\gamma$ contributions to $K_{L}%
\rightarrow\mu^{+}\mu^{-}$\cite{DummPich,IsidoriU03}.\smallskip

\item[--] The $K_{S}\rightarrow\pi^{0}\gamma\gamma$ rate is enhanced by
$\eta_{0}$ effects for most values of $G_{8}^{s}/G_{8}$. In addition, the fact
that neither the shape of the $m_{\gamma\gamma}$ spectrum nor the rate for low
$m_{\gamma\gamma}$ are affected by $G_{8}^{s}$ could be exploited to
disentangle the NLO effects (like vector resonances) from the $\eta_{0}$ ones,
and to achieve a clean extraction of $G_{8}^{s}/G_{8}$.

\item[--] In the case of $K^{+}\rightarrow\pi^{+}\gamma\gamma$, the pole
contribution to the total rate is negligible with respect to loops for
$G_{8}^{s}/G_{8}<0$, but can be as large as $20\%$ for $G_{8}^{s}/G_{8}>0$
(poles are then needed to extract $\hat{c}$). This situation arises from the
fact that the $G_{8}$, $G_{8}^{s}$ and $G_{27}$ pole contributions are of the
same order. At the low energy end of the $\gamma\gamma$ spectrum, the recent
experimental upper bound \cite{E949} points towards $G_{8}^{s}/G_{8}<0$
(still, theoretical uncertainties are large in this small corner of phase-space).

\item[--] The effects of $\eta,\eta^{\prime}$ and $G_{8}^{s}$ on
$K_{L}\rightarrow\pi^{0}\pi^{0}\gamma\gamma$ have proved negligible.\smallskip

\item[--] Relying on Ref.\cite{DAmbrosioP97Klppg}, the $K_{L}\rightarrow
\pi^{+}\pi^{-}\gamma$ data are found compatible with a large range of
$G_{8}^{s}/G_{8}$ values. A precise extraction seems beyond reach given the
many unknown phenomenological parameters.\smallskip

\item[--] The pole contributions to $\Delta M_{LS}^{LD}$ were also briefly
discussed. As for $K_{L}\rightarrow\gamma\gamma$, they arise from $Q_{8}^{s}$
and $Q_{27}$ (i.e., $\hat{Q}_{1}^{2}$). Yet, this time, non-negligible $Q_{8}$
(i.e., $\hat{Q}_{6}^{2}$) effects are expected at NLO from the presence of
$K\rightarrow PP\rightarrow\bar{K}$ loops. The poles result in a negative
contribution to $\Delta M_{LS}$, while loop and short-distance $W$-box diagram
contributions are positive. Conservative bounds on the latter two lead to
$|G_{8}^{s}/G_{8}|<0.6$. For $G_{8}^{s}/G_{8}\simeq-1/3$, large cancellations
are expected between the loop and pole long-distance contributions.\smallskip\bigskip
\end{itemize}

\textit{QCD-inspired operator basis and the penguin contribution to the
}$\Delta I=1/2$\textit{\ rule}

\begin{itemize}
\item[--] The three weak operators $(Q_{8},Q_{8}^{s},Q_{27})$ of $U(3)$ ChPT
can be related to the low-energy realizations of the current-current and
penguin operators $(\hat{Q}_{1},\hat{Q}_{2},\hat{Q}_{6})$. QCD-inspired
theoretical expectations then lead to the range $\left(  G_{8}^{s}%
/G_{8}\right)  _{th}=-0.38\pm0.12$, compatible with all the phenomenological
constraints given above. The non-perturbative evolution of current-current
operators is therefore rather smooth, while the $\hat{Q}_{6}$ penguin is
significantly enhanced, and responsible for about 2/3 of the $\Delta I=1/2$
rule at the hadronic scale. This large penguin fraction is compatible with the
small $\varepsilon^{\prime}/\varepsilon$ observed, though a significant
cancellation with isospin breaking effects might be welcome.\smallskip

\item[--] For $K_{L}\rightarrow\gamma\gamma$, $K_{L}\rightarrow\pi^{+}\pi
^{-}\gamma$ and $\Delta M_{LS}^{LD}$, the well-known vanishing of pole
contributions at lowest order in $SU(3)$ ChPT and the subsequent pathological
sensitivity to NLO effects are explained. Indeed, the transitions
$K_{L}\rightarrow\bar{u}u\rightarrow\gamma\gamma/\pi^{+}\pi^{-}\gamma/K_{L}$,
i.e. through $Q_{1}$, cannot be caught in $SU(3)$ at LO simply because there
are not enough independent weak operators. This does not occur in $U(3)$,
where the $\bar{u}u$ leading order effect can be identified. Contributions
from NLO (in particular from the $\bar{d}d,\bar{s}s$ transitions driven by
$Q_{6}$) should then behave according to chiral counting, except for $\Delta
M_{LS}^{LD}$ where $K\rightarrow PP\rightarrow\bar{K}$ loops are present.\pagebreak 
\end{itemize}

\textit{Theoretical progress in the treatment of pole amplitudes}

\begin{itemize}
\item[--] We have proposed a two-step procedure to deal with pole amplitudes:
first, the identification of vanishing contributions through the enforcement
of theoretical masses (i.e., working consistently at a given order), then the
restoration of correct analytical properties by setting the poles at their
right places for the remaining contributions only.\smallskip

\item[--] The weak mass term $Q_{8}^{m}$ has been shown to disappear in all
the modes considered, in both $SU(3)$ and $U(3)$ ChPT, as long as one is
working consistently at a given order (see above). This is non-trivial since
pole amplitudes involve both off-shell weak transitions\cite{BijnensPP98} and
the WZW action.\smallskip

\item[--] The connection between $U(3)$ and $SU(3)$ ChPT has been analyzed in
details for all modes and, though sometimes large, $\eta_{0}$ effects are not
incompatible with naive $SU(3)$ ChPT power counting. Indeed, they can be
reproduced, to a good approximation, with four $SU(3)$ NLO counterterms
saturated by $\eta_{0}$ exchanges. For $K_{L}\rightarrow\gamma\gamma$, a
complete analysis at $\mathcal{O}(p^{6})$ has been presented, showing in
particular that decay constant corrections do not occur, and that saturating
the $K_{L}\rightarrow\gamma\gamma$ amplitude with $\hat{Q}_{1}$ amounts
essentially to assume that a combination of scalar-dominated counterterms is small.\bigskip
\end{itemize}

In conclusion, a consistent picture seems to emerge from our analysis. In
particular, naive expectations from QCD are already well-supported by
phenomenological constraints. Additional experimental information from
radiative $K$ decays are eagerly awaited, as they could give further insight
into the interplay between strong and weak interactions at low energies.

\vskip  1 true cm

\textbf{Acknowledgments:} C.S. and S.T. are pleased to thank Gino Isidori for
numerous discussions and encouragements. Also, they acknowledge the many
useful interactions with participants of the LNF Spring School, LNF Spring
Institute and Kaon 2005 International Workshop, where parts of this work were
presented. J.-M.\thinspace G. thanks Gino Isidori and Giulia Pancheri for
their kind hospitality during his visit at the Laboratori Nazionali di
Frascati. Many thanks also to Ulrich Nierste for sharing with us his updated
value of $\Delta M_{LS}^{SD}$.

C.S. and S.T. are supported by IHP-RTN, EC contract No. HPRN-CT-2002-00311
(EURIDICE). J.-M. G. acknowledges support by the Belgian Federal Office for
Scientific, Technical and Cultural Affairs through the Interuniversity
Attraction Pole P5/27.

\appendix  

\section*{Appendix: $K_{L}\rightarrow\gamma\gamma$ in $SU(3)$ ChPT at
$\mathcal{O}(p^{6})$}

In the text, we argued that corrections to the weak vertices $K_{L}%
\rightarrow\pi^{0},\eta_{8}$, to the $\pi^{0},\eta_{8}$ propagators and to the
$\pi^{0},\eta_{8}\rightarrow\gamma\gamma$ vertices (in particular through
$F\rightarrow F_{\pi},F_{\eta8}$) should either reconstruct the dominant
$\bar{s}d(\overline{d}s)\rightarrow\bar{u}u\rightarrow\gamma\gamma$
transition, or vanish (to a large extent). This is a strong statement since it
implies a very specific interplay between the weak and strong sectors of
$SU(3)$ ChPT, and between $\mathcal{O}(p^{4})$ and $\mathcal{O}(p^{6})$
counterterms. It is the purpose of this appendix to check this assertion by
performing the full calculation of $K_{L}\rightarrow\gamma\gamma$ in $SU(3)$
ChPT at $\mathcal{O}(p^{6})$. In particular, we will detail all the
cancellations occurring between the various corrections depicted in Fig.16.

\subsection*{Full $\mathcal{O}(p^{6})$ amplitude}

From the diagrams of Fig.16, summing over the $\pi^{0}$ and $\eta_{8}$ poles,
we obtain:%
\begin{equation}
\mathcal{A}^{\mu\nu}\left(  K_{L}\rightarrow\gamma\gamma\right)
=\frac{m_{K}^{2}\alpha}{8\pi^{3}F}\left[  \mathcal{F}\right]  i\varepsilon
^{\mu\nu\rho\sigma}k_{1\rho}k_{2\sigma}\;, \label{Ap1}%
\end{equation}
with%
\begin{align}
\mathcal{F}  &  =-\left(  G_{8}+\frac{2}{3}G_{27}\right)  \frac{A_{K}-A_{\pi}%
}{m_{K}^{2}-m_{\pi}^{2}}\nonumber\\
&  -\,\,G_{8}\left[  1024\pi^{2}\left(  2L_{7}+L_{8}+\frac{L_{8}^{(6)}%
+6L_{9}^{(6)}}{9}+\frac{2N_{12}+2N_{13}-N_{6}}{8}+\frac{N_{1}^{(6)}}{12}%
+N_{i}^{(6)}\right)  \right] \nonumber\\
&  +G_{27}\left[  1024\pi^{2}\left(  2L_{7}+L_{8}+\frac{L_{8}^{(6)}%
+6L_{9}^{(6)}}{9}+D_{i}+D_{i}^{(6)}\right)  \right] \nonumber\\
&  +\,G_{8}^{m}\left[  1024\pi^{2}\left(  2L_{7}+L_{8}+\frac{L_{8}%
^{(6)}+6L_{9}^{(6)}}{9}\right)  \right]  \;, \label{Ap2}%
\end{align}
$A_{i}=m_{i}^{2}(D_{\varepsilon}-\log m_{i}^{2}/\mu^{2}+1)$ and
$D_{\varepsilon}=\frac{2}{\varepsilon}-\gamma+\log4\pi$. We have used the
standard basis of Gasser-Leutwyler\cite{GasserL84} for the $\mathcal{O}%
(p^{4})$ strong counterterms ($L_{i}$) and the
Ecker-Kambor-Wyler\cite{EckerKW93} one for the $\mathcal{O}(p^{4})$ octet
$\Delta S=1$ counterterms ($N_{i}$). In the third diagram, the CT
(counterterm) contribution originates from the odd-parity $\mathcal{O}(p^{6})$
strong Lagrangian, which is written in the
Ebertshauser-Fearing-Scherer\cite{EbertshauserFS02} or
Bijnens-Girlanda-Talavera\cite{BijnensGT02} basis (up to an overall
normalization):%
\begin{align}
\mathcal{O}_{8}^{EFS}  &  =\mathcal{O}_{7}^{BGT}=\frac{1}{\left(  4\pi
F\right)  ^{2}}L_{8}^{(6)}i\varepsilon_{\mu\nu\alpha\beta}\langle\chi_{-}%
f_{+}^{\mu\nu}f_{+}^{\alpha\beta}\rangle\,,\label{Ap3}\\
\mathcal{O}_{9}^{EFS}  &  =\mathcal{O}_{8}^{BGT}=\frac{1}{\left(  4\pi
F\right)  ^{2}}L_{9}^{(6)}i\varepsilon_{\mu\nu\alpha\beta}\langle\chi
_{-}\rangle\langle f_{+}^{\mu\nu}f_{+}^{\alpha\beta}\rangle\,, \label{Ap4}%
\end{align}
with $\chi_{\pm}\equiv U\chi^{\dagger}\pm\chi U^{\dagger}$, $f_{\pm}^{\mu\nu
}\equiv F_{L}^{\mu\nu}\pm UF_{R}^{\mu\nu}U^{\dagger}$ (in our case $\chi=rM$
with $M$ the light quark mass matrix, $F_{L}^{\mu\nu}=F_{R}^{\mu\nu}%
=-eQF^{\mu\nu}$ with $F^{\mu\nu}$ the QED field strength tensor and $Q$ the
light quark charge matrix). In the last diagram, the CT contribution
originates from the odd-parity $\mathcal{O}(p^{6})$ weak Lagrangian. To our
knowledge, no operator basis has been set up for it yet. For the purpose of
the present analysis, only one of the direct $K_{L}\rightarrow\gamma\gamma$
couplings will be needed explicitly:%
\begin{equation}
\mathcal{O}_{\left|  \Delta S\right|  =1}^{(6)}=\frac{G_{8}}{2\left(
4\pi\right)  ^{2}}N_{1}^{(6)}\,i\varepsilon_{\mu\nu\alpha\beta}\langle
\lambda_{6}\chi_{-}\rangle\langle f_{+}^{\mu\nu}f_{+}^{\alpha\beta}\rangle\;.
\label{Ap5}%
\end{equation}
All the others are collectively denoted by $N_{i}^{(6)}$ (octet part). Note
that the $\mathcal{O}(p^{4})$ CTs $L_{i}$ and $N_{i}$ alone suffice to absorb
the $G_{8}$ part of the loop divergence, $L_{i}^{(6)}$ and $N_{i}^{(6)}$ being
thus separately finite. Finally, also the 27-plet CTs, though needed for
renormalization, will not be specified explicitly, and are collectively
denoted by $D_{i}$ \cite{Esposito91,KamborMW90} and $D_{i}^{(6)} $.

Note also that, to reach Eq.(\ref{Ap2}), the initial $K_{L}$ mass $M_{K}^{2}$
has been expressed back in terms of the bare mass $m_{K}^{2}$ (see
Ref.\cite{GasserL84}). Alternatively, the weak rotation can be performed to
discard the $K_{L}\rightarrow\pi^{0},\eta_{8}$ $\mathcal{O}\left(
p^{2}\right)  $ vertices\cite{EckerPR87,EckerPR88}.%
\begin{figure}
[t]
\begin{center}
\includegraphics[
height=0.9132in,
width=5.8574in
]%
{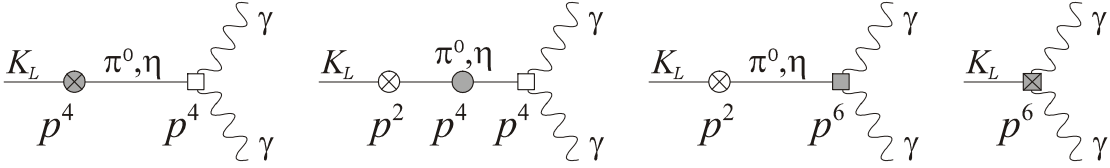}%
\caption{$K_{L}\rightarrow\gamma\gamma$ at $\mathcal{O}\left(  p^{6}\right)
$. Grey vertices stand for local counterterms and meson loops. }%
\end{center}
\end{figure}

\subsection*{Discussion of the counterterms}

A first encouraging observation is that many counterterms drop out when
summing over the $\pi^{0}$ and $\eta_{8}$ poles ($N_{5,8,10,11}$ and
$L_{4,5,6}$) or taking the external particles on-shell (like $N_{29}$,$N_{31}%
$), while the final combination of chiral logs is very simple and very small
(it vanishes for $\mu\sim550$ MeV). In particular, the a priori significant
correction due to $SU(3)$ breaking effects in the decay constants (i.e.,
$L_{4,5}$) drops out, the impact of $F_{\eta8}/F_{\pi}\neq1$ being thus at
least of $\mathcal{O}(p^{8})$. But this is not the end of the story: there are
still many interplays between the counterterms, though it is necessary to go
beyond the strict $SU(3)$ framework to get hold of them.

As is well-known \cite{GasserL84}, $L_{7}$ is well-described by a tree-level
$\eta_{0}$ exchange, which gives it an abnormal counting $L_{7}\sim N_{c}^{2}$
(still, see the discussion in \cite{PerisR94}). In the same way, $N_{13}$,
$L_{9}^{(6)}$ and $N_{1}^{(6)}$ (which also have zero anomalous dimensions)
can be saturated by the $\eta_{0}$ resonance, surviving then as well in the
large-$N_{c}$ limit. The explicit reductions of the $U(3)$ weak operators to
$SU(3)$ read ($\bar{\eta}_{0}\equiv\sqrt{2/3}\eta_{0}/F$)%
\begin{align}
Q_{8}  &  =4\left[  \langle\lambda_{6}L_{\mu}L^{\mu}\rangle-F^{2}%
\langle\lambda_{6}L_{\mu}\rangle\left(  D^{\mu}\bar{\eta}_{0}\right)  \right]
\,,\nonumber\\
Q_{8}^{s}  &  =-6F^{2}\langle\lambda_{6}L_{\mu}\rangle\left(  D^{\mu}\bar
{\eta}_{0}\right)  \,,\label{Ap6a}\\
Q_{8}^{m}  &  =F^{4}\left[  \cos\left(  \bar{\eta}_{0}\right)  \langle
\lambda_{6}\chi_{+}\rangle+i\sin\left(  \bar{\eta}_{0}\right)  \langle
\lambda_{6}\chi_{-}\rangle\right]  \,,\nonumber
\end{align}
where the left-hand side $\in U(3)$ and the right-hand side $\in SU(3)$
($Q_{27}$ does not couple to the singlet). Similarly, the WZW $U(3)$
Lagrangian is reduced as \cite{KaiserL00}%
\begin{equation}
\mathcal{L}_{WZW}^{U(3)}=\mathcal{L}_{WZW}^{SU(3)}+\frac{iN_{c}}{288\pi^{2}%
}\varepsilon_{\mu\nu\alpha\beta}\left[  \frac{3}{4}\langle f_{+}^{\mu\nu}%
f_{+}^{\alpha\beta}\rangle+\frac{1}{4}\langle f_{-}^{\mu\nu}f_{-}^{\alpha
\beta}\rangle+i\langle D^{\mu}UD^{\nu}U^{\dagger}f_{+}^{\alpha\beta}%
\rangle\right]  \left\langle i\bar{\eta}_{0}\right\rangle \,. \label{Ap6b}%
\end{equation}
A straightforward computation then gives%
\begin{align}
L_{7}  &  =-\frac{N_{c}F^{2}}{144m_{0}^{2}},\;\;\;\;\;N_{13}=\frac{N_{c}F^{2}%
}{18m_{0}^{2}}\left(  1+\frac{3}{2}\frac{G_{8}^{s}}{G_{8}}-\frac{G_{8}^{m}%
}{G_{8}}\right)  \;,\nonumber\\
L_{9}^{(6)}  &  =-\frac{N_{c}^{2}F^{2}}{144m_{0}^{2}},\;\;\;\;\;N_{1}%
^{(6)}=\frac{N_{c}^{2}F^{2}}{18m_{0}^{2}}\left(  1+\frac{3}{2}\frac{G_{8}^{s}%
}{G_{8}}-\frac{G_{8}^{m}}{G_{8}}\right)  \; \label{Ap7}%
\end{align}
after partial integration and use of the $SU(3)$ equations of motion. Note
that other counterterms (like $N_{24}$, $L_{7}^{(6)}$, $L_{11}^{(6)}$,...)
could be saturated by the $\eta_{0}$, but none contributing for on-shell photons.

Besides, we will set $L_{8}^{(6)}=0$, as suggested by the success of the LO
description of $\pi^{0}\rightarrow\gamma\gamma$. There is then only the
$L_{9}^{(6)}$ counterterm for $\eta\rightarrow\gamma\gamma$ at $\mathcal{O}%
(p^{6})$ (chiral loops and wavefunction renormalization amount to
$F\rightarrow F_{\eta8}$, see Ref.\cite{Pgg}). Using its $\eta_{0}$-saturated
value with $m_{0}\simeq850$ MeV, we obtain $\Gamma\left(  \eta\rightarrow
\gamma\gamma\right)  \simeq0.4\,$keV, instead of $0.1\,$keV without
$L_{9}^{(6)}$. Experimentally, $\Gamma\left(  \eta\rightarrow\gamma
\gamma\right)  =\left(  0.510\pm0.026\right)  $\thinspace keV \cite{PDG04},
which means that $L_{9}^{(6)}$ indeed accounts for the bulk of NLO effects.
This is simply a manifestation of the fact that $U(3)$ at LO reproduces
$\eta,\eta^{\prime}\rightarrow\gamma\gamma$ reasonably well without extra
operators like Eq.(\ref{WZW3}).

Inserting the values (\ref{Ap7}) in Eq.(\ref{Ap2}), we find that the $G_{8}$
and $G_{8}^{m}$ pieces of the $\eta_{0}$-saturated CTs indeed cancel out,
leaving only a term proportional to $G_{8}^{s}+\frac{2}{3}G_{27}$:%
\begin{align}
\mathcal{F}  &  =-128\pi^{2}\left(  G_{8}^{s}+\frac{2}{3}G_{27}\right)
\frac{F^{2}}{m_{0}^{2}}+\text{ chiral\thinspace\thinspace logs}\nonumber\\
&  -1024\pi^{2}\left[  G_{8}\left(  L_{8}+\frac{2N_{12}-N_{6}}{8}+N_{i}%
^{(6)}\right)  -G_{27}\left(  L_{8}+D_{i}+D_{i}^{(6)}\right)  -G_{8}%
^{m}\left(  L_{8}\right)  \right]  \,. \label{Ap9}%
\end{align}
Note that the use of the $U(3)$ lowest order constraint $F=F_{\pi}=F_{\eta0}$
does not affect the observed cancellation since any deviation of $F_{\eta
0}/F_{\pi}$ from unity can be accounted for by varying $m_{0}^{2}$ (saturation
from tree-level exchanges).

The first term is of course exactly the $U(3)$ result in the limit
$m_{0}\rightarrow\infty$ and corresponds to the $\hat{Q}_{1}$ contribution,
i.e. to the transition $\bar{s}d(\overline{d}s)\rightarrow\bar{u}u\rightarrow
\gamma\gamma$. The $G_{27}$ piece ($D_{i},D_{i}^{(6)}$) is suppressed by the
$\Delta I=1/2$ rule and can be discarded, along with the small chiral logs.
The remaining $\mathcal{O}(p^{4})$ counterterms $N_{6},N_{12}$ and $L_{8}$ can
in principle give a large contribution to $K_{L}\rightarrow\gamma\gamma$
because of their large numerical coefficients. Still, they all arise from
scalar exchanges\cite{EckerGPR,EckerKW93} and it is reasonable to assume that
they will behave similarly to the counterterms saturated by pseudoscalar
exchanges, i.e. cancel among themselves to a large extent, or act as a
correction to the leading $\hat{Q}_{1}$ piece. As a further clue, note that
there must be some interplay between $L_{8}$ and $N_{6,12}$ in order to absorb
the contribution of $G_{8}^{m}$ in the $N_{i}$'s. Finally, the fate of the
remaining $\mathcal{O}(p^{6})$ local $K_{L}\rightarrow\gamma\gamma$
counterterms ($N_{i}^{(6)}$) is not clear, but it is at least possible that
they are small.

In conclusion, we have pin-pointed the combination of counterterms on which we
make an assumption when we saturate the $K_{L}\rightarrow\gamma\gamma$ rate by
the $\hat{Q}_{1}$ contribution. We hope that some further work will be able to
get at least an upper bound on this combination, so that the precise
measurements of $K_{L}\rightarrow\gamma\gamma$ (\ref{KL1}) will reliably fix
$G_{8}^{s}/G_{8}$. For now, as explained in the text, there are other
theoretical and phenomenological indications that $G_{8}^{s}/G_{8}$ has a
value around $-1/3$, which then unambiguously implies that the remaining CTs
in Eq.(\ref{Ap9}) must at least partially cancel among themselves.

\subsection*{Comments on $\Delta M_{LS}$}

Finally, let us make a few comments on the similar computation for $\Delta
M_{LS}^{LD}$. As explained in the text, in $SU(3)$ at $\mathcal{O}(p^{4})$,
there are both corrections to the pole amplitudes, similar to the
$K_{L}\rightarrow\gamma\gamma$ ones given in Eq.(\ref{Ap2}), and pseudoscalar
loops like $K\rightarrow PP\rightarrow\bar{K}$. Large cancellations among
corrections are again observed, in particular among those for the decay
constants and masses. It should be noted also that though the terms in
$G_{8}^{m}G_{8,27}$ and ($G_{8}^{m}$)$^{2}$ are non-local for $M(K_{L})$ and
$M(K_{S})$ (i.e., contain some chiral logs), they appear only locally for
$\Delta M_{LS}^{LD}$ (i.e., they are proportional to $\mathcal{O}(p^{4})$
counterterms and can thus be absorbed in the $N_{i}$'s). As at $\mathcal{O}%
(p^{2})$, only the mass difference is unambiguous in Chiral Perturbation
Theory (see Sec.7.1).

Concerning the reconstruction of the $\hat{Q}_{1}$ contribution, the role of
$L_{9}^{(6)}$ and $N_{1}^{(6)}$ is taken up by the $\Delta S=2$ counterterm
$\langle\lambda_{6}\chi_{-}\rangle^{2}$, saturated by the $\eta_{0}$. Again,
$L_{7}$, $N_{13}$ and this $\Delta S=2$ counterterm appear initially with a
complicated coefficient involving $G_{8}^{2}$, but they conspire to
reconstruct the $s\bar{d}\rightarrow\bar{u}u\rightarrow\bar{s}d$ transition,
i.e. a term with the coefficient $(G_{8}^{s}+2G_{27}/3)^{2}$. At this stage,
an expression similar to Eq.(\ref{Ap9}) is reached, but with in addition some
$K\rightarrow PP\rightarrow\bar{K}$ loop functions (and divergences), and a
correspondingly different combination of the remaining counterterms.

In conclusion, because of the presence of $K\rightarrow PP\rightarrow\bar{K} $
loops and $\Delta S=2$ counterterms, it does not appear possible to get extra
information from $\Delta M_{LS}^{LD}$ on the counterterms remaining for
$K_{L}\rightarrow\gamma\gamma$ in Eq.(\ref{Ap9}). Still, the interplays among
corrections leading to the reconstruction of the $\hat{Q}_{1}$ contribution
are found to occur also for $\Delta M_{LS}$.

\end{document}